\documentclass[review]{elsarticle}

\usepackage[cmex10]{amsmath}
\usepackage{threeparttable}
\usepackage{amssymb}
\usepackage{graphicx}
\usepackage{lineno, hyperref}
\usepackage{verbatim}
\usepackage{bm}
\usepackage{multirow,textcomp,booktabs}
\usepackage[usenames,dvipsnames]{color}
\usepackage{colortbl}
\definecolor{mygray}{gray}{.6}
\definecolor{1mygray1}{gray}{.7}
\definecolor{1mygray2}{gray}{.72}
\definecolor{1mygray3}{gray}{.8}
\definecolor{1mygray4}{gray}{.82}
\definecolor{1mygray5}{gray}{.87}
\definecolor{1mygray6}{gray}{.90}

\definecolor{2mygray1}{gray}{.7}
\definecolor{2mygray2}{gray}{.73}
\definecolor{2mygray3}{gray}{.8}
\definecolor{2mygray4}{gray}{.83}
\definecolor{2mygray5}{gray}{.86}
\definecolor{2mygray6}{gray}{.90}

\definecolor{3mygray1}{gray}{.7}
\definecolor{3mygray2}{gray}{.79}
\definecolor{3mygray3}{gray}{.82}
\definecolor{3mygray4}{gray}{.86}
\definecolor{3mygray5}{gray}{.9}

\definecolor{4mygray1}{gray}{.7}
\definecolor{4mygray2}{gray}{.76}
\definecolor{4mygray3}{gray}{.78}
\definecolor{4mygray4}{gray}{.80}
\definecolor{4mygray5}{gray}{.95}

\definecolor{5mygray1}{gray}{.7}
\definecolor{5mygray2}{gray}{.72}
\definecolor{5mygray3}{gray}{.80}
\definecolor{5mygray4}{gray}{.87}
\definecolor{5mygray5}{gray}{.95}

\definecolor{6mygray1}{gray}{.7}
\definecolor{6mygray2}{gray}{.71}
\definecolor{6mygray3}{gray}{.75}
\definecolor{6mygray4}{gray}{.80}
\definecolor{6mygray5}{gray}{.85}
\definecolor{6mygray6}{gray}{.90}
\definecolor{6mygray7}{gray}{.95}

\usepackage{balance}
\usepackage{times,amsmath,epsfig}
\usepackage{subfigure}
\usepackage[round-mode=places, round-integer-to-decimal, round-precision=1]{siunitx}
\usepackage{mathrsfs}
\usepackage{amsmath}
\usepackage{bm}
\usepackage{url}
\usepackage{doi}
\biboptions{numbers,sort&compress}

\journal{Journal of \LaTeX\ Templates}

\begin{document}
\begin{frontmatter}

\title{{New Design Paradigm of Distortion Cost Function for Efficient JPEG Steganography}}

\author[mymainaddress]{Wenkang Su}
\ead{suwk3@mail.sysu.edu.cn}

\author[mymainaddress]{Jiangqun Ni\corref{mycorrespondingauthor}}
\cortext[mycorrespondingauthor]{Corresponding author}
\ead{issjqni@mail.sysu.edu.cn}

\author[mymainaddress]{Xianglei Hu}
\ead{springhuxl@gmail.com}

\author[mysecondaryaddress]{Jiwu Huang}
\ead{jwhuang@szu.edu.cn}

\address[mymainaddress]{School of Computer Science and Engineering, Sun Yat-sen University, Guangzhou, China}
\address[mysecondaryaddress]{College of Information Engineering, Shenzhen University, Shenzhen, China}

\begin{abstract}
Recently, with the introduction of JPEG phase-aware steganalysis features, e.g., GFR, the design of JPEG steganographic distortion cost function turns to maintain not only the statistical undetectability in DCT domain but also in spatial domain. To tackle this issue, this paper presents a novel paradigm for the design of JPEG steganographic distortion cost function, which calculates the distortion cost via a generalized Distortion Cost Domain Transformation (DCDT) function. The proposed function comprises the decompressed pixel block embedding changes and their corresponding embedding distortion costs for unit change, where the pixel embedding distortion costs are represented in a more general exponential model, aiming to flexibly allocate the embedding data. In this way, the JPEG steganography could be formulated as the optimization problem of minimizing the overall distortion cost in its decompressed spatial domain, which is equivalent to maximizing its statistical undetectability against JPEG phase-aware steganalysis features. Experimental results show that the proposed DCDT equipped with HiLL (a spatial steganographic distortion cost function) is superior to other state-of-the-art JPEG steganographic schemes, e.g., UERD, J-UNIWARD, and GUED in resisting the detection of JPEG phase-aware feature-based steganalyzers GFR and SCA-GFR, and rivals BET-HiLL with one order of magnitude lower computational complexity, along with the possibility of being further improved by considering the mutually dependent embedding interactions. In addition, the proposed DCDT is also verified to be effective for different image databases and quality factors.

\end{abstract}
\begin{keyword}
information hiding  \sep JPEG steganography \sep distortion cost function\sep domain transformation \sep exponential model
\end{keyword}

\end{frontmatter}


\section{Introduction}
Steganography is the science and art of covert communication without drawing suspicion from the Warden \cite{history,Benchmarking}. With the rapid development of multimedia information technology, e.g., image, audio, and video, the steganography technology and its applications \cite{ LI2021107920, YU2020107343, QIAO2021108048, ZhangYi, SU20111901, Wang9153904, Yi8626153, VoIP, li2014a, GMRF-Su, guo2014uniform, guo2015using, holub2014universal, GUED, BET}  have also made great progress in the past decades. And among them, the content-adaptive JPEG (image) steganography \cite{guo2014uniform, IUERD, guo2015using, holub2014universal, GUED, BET}, which conceals secret messages in quantized DCT (Discrete Cosine Transform) coefficients, is currently the most popular and practical one since the `jpg' format image is most commonly used in our lives.

With the emergence of the breakthrough coding method -- STCs (Syndrome-Trellis Codes) \cite{filler2011minimizing} for minimal distortion embedding, the majority of the prevailing JPEG steganographic schemes focus on the design of effective steganographic distortion cost function, e.g., UERD \cite{guo2015using}, J-UNIWARD \cite{holub2014universal}, GUED \cite{GUED}, and BET \cite{BET}. To be specific, UERD uses block energy, i.e., the sum of the absolute value of dequantized DCT coefficients within the $8 \times 8$ DCT block, and JPEG quantization step to construct the distortion cost function. And the distortion function in J-UNIWARD is defined as the absolute sum of relative changes of the wavelet coefficients w.r.t. the cover image, where the wavelet coefficients are obtained by filtering the decompressed image using the Daubechies 8-tap wavelet directional filter bank. In consideration of the deficiency in UERD, the GUED is proposed to improve the distortion measures for DCT mode and DCT block, i.e., the absolute sum of decompressed spatial pixel block embedding changes and the absolute sum of Gabor residuals on decompressed spatial pixel block, respectively. To further improve the capability of JPEG steganography against the detection of JPEG phase-aware feature-based steganalyzers, e.g., GFR \cite{song2015steganalysis} and SCA-GFR \cite{denemark2016steganalysis}, BET directly utilizes the embedding entropy of decompressed spatial pixel block to construct DCT block distortion measure, and by which, BET becomes currently the most secure JPEG steganographic scheme.

The success of J-UNIWARD, GUED, and BET against the detection of JPEG phase-aware feature-based steganalyzers indicates that JPEG steganography should maintain not only the statistical undetectability in DCT domain but also in spatial domain. Following this philosophy of distortion cost function design, in this paper, we propose a novel paradigm for JPEG steganography,
namely {\textbf{D}}istortion {\textbf{C}}ost {\textbf{D}}omain {\textbf{T}}ransformation (DCDT) based JPEG steganography scheme, which formulates the JPEG steganography as the optimization on minimizing the overall distortion cost in its
decompressed spatial domain. The basis of our proposed scheme is that the embedding priority for both the $8\times8$ DCT block and its decompressed block in spatial should be the same since they represent the same image information. In our proposed scheme, a generalized distortion cost domain transformation function $f$ is introduced to directly transform the decompressed spatial distortion cost into JPEG domain with the assumption that the spatial distortion cost is linearly proportional to the amplitude of embedding modification in its decompressed spatial domain. To further maintain the statistical undetectability, an exponential model is then developed for spatial distortion cost to improve the construction of $f$. Extensive experiments show that the proposed scheme equipped with HiLL has a more comprehensive security performance improvement than UERD with the same computational complexity, and is superior to J-UNIWARD and GUED in resisting the detection of GFR and SCA-GFR, along with the possibility of being further improved by considering the mutually dependent embedding interactions. Besides, it can also rival the state-of-the-art (SOTA) BET-HiLL with one order of magnitude lower computational complexity. What's more, the proposed scheme is also effective and widely applicable for other image databases and a variety of Quality Factors (QFs).

The remainder of this paper is organized as follows. In the next section, we firstly introduce the basis and motivation behind the proposed scheme in subsection 2.1, and then the selection strategy of spatial steganographic distortion cost function will be discussed in subsection 2.2. Subsequently, the construction of the generalized distortion cost domain transformation function is given in subsection 2.3. Additionally, we further make an extension for the proposed scheme in terms of mutually dependent embedding in section 2.4, which is followed by the extensive experimental results and analysis in section 3. Finally, the paper is concluded in section 4, where we summarize the most important contributions given in this paper.

\section{The proposed novel paradigm for the design of JPEG steganographic distortion cost function}
In this section, we propose a novel paradigm for the design of JPEG steganographic distortion cost function, which obtains the JPEG distortion cost via directly transforming the spatial embedding distortion cost into JPEG domain. In the following, the basis and motivation behind this proposed scheme will be firstly elaborated. And then, the selection of spatial steganographic distortion cost function for the proposed scheme will be discussed subsequently. Next, the construction of the proposed generalized distortion cost domain transformation function, which is the core of our proposed scheme, will be explained in detail. Finally, the extension to mutually dependent embedding for the proposed scheme will be further presented.

\subsection{The basis and motivation behind the proposed scheme}
Concerning the JPEG steganography, it is well known that when we modify the DCT coefficient $x_{a,b}^{m,n}$, i.e., the one at mode $(a,b)$ in the ${(m,n)^{th}}$ DCT block, the corresponding spatial embedding changes can be easily derived by its inverse DCT transformation. Since JPEG compression is based on block DCT transformation, then the decompressed spatial embedding changes would only happen within its corresponding $8 \times 8$ pixel block, which is associated with the quantization step $q_{a,b}$, irrespective of image content. Thus, the relationship between the DCT domain embedding modification and the spatial embedding changes can be explicitly expressed as:
\begin{equation}\label{eq:J2S}
{{\bf{s}}_{a,b}} = {q_{a,b}} \cdot ({{\bf{A}}^T} * {{\bf{t}}_{a,b}} * \bf{A}),
\end{equation}
where
\begin{equation}\label{eq:J2S-2}\bf{A} = \left[ {\begin{array}{*{20}{c}}
a&a&a&a&a&a&a&a\\
b&d&e&g&{ - g}&{ - e}&{ - d}&{ - b}\\
c&f&{ - f}&{ - c}&{ - c}&{ - f}&f&c\\
d&{ - g}&{ - b}&{ - e}&e&b&g&{ - d}\\
a&{ - a}&{ - a}&a&a&{ - a}&{ - a}&a\\
e&{ - b}&g&d&{ - d}&{ - g}&b&{ - e}\\
f&{ - c}&c&{ - f}&{ - f}&c&{ - c}&f\\
g&{ - e}&d&{ - b}&b&{ - d}&e&{ - g}
\end{array}} \right],
\end{equation}
\begin{equation}\label{eq:J2S-3}
{\rm{      }}\left[ {\begin{array}{*{20}{c}}
a\\
b\\
c\\
d\\
e\\
f\\
g
\end{array}} \right] = \frac{1}{2}\left[ {\begin{array}{*{20}{c}}
{\cos \left( {\pi /4} \right)}\\
{\cos \left( {\pi /16} \right)}\\
{\cos \left( {\pi /8} \right)}\\
{\cos \left( {3\pi /16} \right)}\\
{\cos \left( {5\pi /16} \right)}\\
{\cos \left( {3\pi /8} \right)}\\
{\cos \left( {7\pi /16} \right)}
\end{array}} \right],
\end{equation}
`$*$' indicates the matrix multiplication, and ${{\bf{A}}^T}$ is the transpose of {\bf{A}}, ${\bf{t}}_{a,b}$ represents the solitary modification on mode $(a,b)$ among the 64 DCT modes, ${{\bf{s}}_{a,b}}$ denotes the resultant corresponding spatial ${8 \times 8}$ pixel block embedding changes.

As we know, the JPEG compression is based on block-wise DCT transformation, so the $8\times8$ DCT block represents the same information with its corresponding decompressed $8\times8$ pixel block, then the embedding priority of the $8\times8$ DCT and pixel block shall be the same, which in turn constitutes the basis of our proposed scheme. In addition, the objective of content-adaptive spatial steganography is to minimize their overall distortion for given payload under the framework of minimal distortion embedding \cite{filler2011minimizing}. Therefore, referring to Eq. \eqref{eq:J2S}, if we can measure the spatial distortion for arbitrary modification amplitude, then the overall additive distortion of JPEG steganography in its decompressed spatial domain can be accordingly obtained. As thus, we can formulate the JPEG steganography under the framework of minimal distortion embedding as the optimization on minimizing the overall distortion cost in its decompressed spatial domain, and therefore to improve the performance of JPEG steganography by maintaining the statistical undetectability in both spatial and DCT domains.

\subsection{Discussion on the selection of spatial steganographic distortion cost function}
As the key part in calculating the overall distortion cost in the decompressed spatial domain, the selection of spatial steganographic distortion cost function for the proposed scheme is of vital importance. With regard to the method of calculating the spatial distortion cost, there are many candidates, such as WOW \cite{WOW}, S-UNIWARD \cite{holub2014universal}, HiLL \cite{li2014a}, MiPOD \cite{sedighi2016content} and etc. As stated earlier, the $8\times8$ DCT and pixel block has the same embedding priority, then the better the spatial steganography cost function is utilized, the higher the security of the proposed scheme should be. Notably, the HiLL would be an excellent candidate because of its excellent security performance and minimal computational complexity\footnote{Actually, we have also tested other spatial steganographic distortion cost functions in section 3.3, and find that HiLL is indeed the one which yields the best security performance.}. To analyze its feasibility, we then make a simple experiment in the following, i.e., calculating and comparing the similarity in evaluating the DCT block embedding priority between HiLL and other JPEG steganographic distortion cost functions.

Without loss of generality, we randomly select 2,000 cover images from BOSSBase ver1.01 \cite{bas2011break} at Q75 and Q95{\footnote{In the rest of this paper, for brevity, we represent QF=75 and QF=95 by Q75 and Q95, respectively.}} separately and then use UERD, J-UNIWARD, GUED, and HiLL to calculate the embedding cost for each $8 \times 8$ DCT or decompressed pixel block within the cover. In our experiment, it should be noted that the block embedding cost with HiLL is expressed by the sum of 64 pixels' embedding costs within this block of the decompressed image, while for J-UNIWARD, it is expressed by the reciprocal sum of the absolute value of $23\times23$ wavelet filter residuals w.r.t this block in three directions. In addition, we will also randomly generate a set of DCT block embedding costs denoted as \emph{Rand}, as a comparison to verify the validity of this experiment. Since the block embedding priority is determined by the block embedding cost, thus, the similarities of (DCT or pixel) block embedding priority among different steganographic schemes can be evaluated by calculating the similarities of their block embedding costs. As regards the choice of metric for similarity, we adopt the \emph{Spearman Correlation Coefficient} (\emph{SCC}) \cite{CORR_COEF}, which is one of the three popular statistical correlation coefficients and corresponds to the `corr' Matlab command with type `Spearman'. The sign `+' and `-' of \emph{SCC} represent positive correlation and negative correlation, respectively, and the magnitude represents the degree of correlation (0 is irrelevant, 1 is completely linear relevant). Finally, the average \emph{SCCs} over 2000 cover images at Q75 and Q95 are summarized in Table \ref{tab:SCCs}.

\begin{table} [thbp]
\renewcommand\arraystretch{1}
\centering
\caption{The average \emph{Spearman Correlation Coefficients} (\emph{SCCs}) over 2000 cover images at Q75 and Q95 between the DCT block embedding cost (Rand, UERD, J-UNIWARD, GUED) and the pixel block embedding cost (HiLL), respectively.}\label{tab:SCCs}
\begin{tabular}{ccc}
  \toprule[1pt]
    \centering
    {Different schemes} &Q75 &Q95\\
    \midrule[0.8pt]
    \emph{SCC}(\emph{Rand,HiLL}) 	&$0.0$  &$0.0$ \\
    \midrule[0.8pt]
    \emph{SCC}(\emph{UERD,HiLL}) 	&$+0.7801$   &$+0.7777$  \\
    \midrule[0.8pt]
    \emph{SCC}(\emph{J-UNIWARD,HiLL}) 	&$+0.8420$  &$+0.8584$  \\
   \midrule[0.8pt]
    \emph{SCC}(\emph{GUED,HiLL}) 	&$+0.8666$  &$+0.8962$  \\
    \bottomrule[1pt]
\end{tabular}
\end{table}

Referring to the results in Table \ref{tab:SCCs}, it is observed that \emph{SCC}(\emph{Rand,HiLL}) is close to 0, while others are around 0.8. Since the block embedding cost with Rand is randomly generated, while for UERD, J-UNIWARD, GUED, and HiLL, they are all well designed based on the statistical characteristics of cover image, so this result indicates that the proposed similarity metric \emph{SCC} is reasonable. In addition, comparing \emph{SCC}(\emph{J-UNIWARD,HiLL}) with \emph{SCC}(\emph{UERD,HiLL}), it is observed that J-UNIWARD is closer to HiLL than UERD in evaluating the block embedding priority along with higher security performance against steganalyzers, e.g., GFR. It is the same for GUED and J-UNIWARD. Furthermore, reviewing the performance of BET \cite{BET} and GUED \cite{GUED}, it shows that the BET-HiLL whose block embedding cost is constructed from HiLL is also superior to GUED in resisting the detection of GFR. In this regard, it is convinced that if the evaluation of block embedding priority of a JPEG steganographic scheme is closer to HiLL's, then it would be more secure. Therefore, if the DCT block embedding priority for a JPEG steganographic scheme is evaluated with HiLL on the corresponding block of the decompressed image, better security performance is expected to be achieved.

\subsection{Construction of the proposed distortion cost domain transformation function}
Referring to section 2.1, we know that the spatial distortion cost for arbitrary modification amplitude should be defined when we intend to formulate the JPEG steganography as the optimization on minimizing the overall distortion cost in its decompressed spatial domain. Note that the unit modification ($+1/-1$) on DCT coefficient will lead to non-unit spatial embedding changes, and on the other hand, although there exist a variety of fairly good distortion functions in spatial domain, they are almost all designed for measuring the distortion on unit embedding change.
In this regard, we make a simple yet effective assumption that the spatial distortion cost is linearly proportional to the amplitude of modification for a pixel. As thus, for the $\pm1$ modification on DCT coefficient $x_{a,b}^{m,n}$, the resulting spatial additive distortion can be expressed as:
\begin{equation}\label{eq:rho}
\rho _{a,b}^{m,n} = \sum\limits_{i = 1}^8 {\sum\limits_{j = 1}^8 {d_{m,n}{(i,j)} \cdot \left| {s_{a,b}{(i,j)}} \right|} },
\end{equation}
where $d_{m,n}{(i,j)}$ represents the spatial distortion cost of the ${(i,j)^{th}}$ pixel in corresponding block of decompressed image for unit embedding change, $\left| {s_{a,b}{(i,j)}} \right|$ indicates the resulting spatial embedding changes within the corresponding $8\times8$ block due to the unit embedding modification at DCT mode $(a,b)$, which can be obtained by Eq. \eqref{eq:J2S}. By taking into account the statistics both in spatial and DCT domains, the $\rho _{a,b}^{m,n}$ in Eq. \eqref{eq:rho} could well evaluate the resulting distortions in both spatial and DCT domains arising from the embedding modification at DCT coefficient $x_{a,b}^{m,n}$, and be adopted as the distortion cost for the proposed JPEG steganographic scheme. Since $\rho _{a,b}^{m,n}$ is obtained by transforming the spatial distortion cost into DCT domain, the proposed scheme can then be referred to as {\bf{D}}istortion {\bf{C}}ost {\bf{D}}omain {\bf{T}}ransformation (DCDT) based JPEG steganographic scheme, and the Eq. \eqref{eq:rho} can be formulated as a distortion cost domain transformation function $f\left( {{{\mathop{\bf{d}}\nolimits} _{m,n}},{{\mathop{\bf{s}}\nolimits}_{a,b}}} \right)$ as well, where ${\mathop{\bf{d}}\nolimits} _{m,n}$ and ${\mathop{\bf{s}}\nolimits}_{a,b}$ are the ${(m,n)^{th}}$ decompressed $8\times8$ pixel block distortion costs and decompressed $8\times8$ spatial embedding changes for solitary modification on DCT mode $(a,b)$, respectively.

\begin{figure*}[h]
\centering
{
\subfigure[]{\label{fig:J-UNIWARD_Q75}\includegraphics[width=0.42\textwidth]{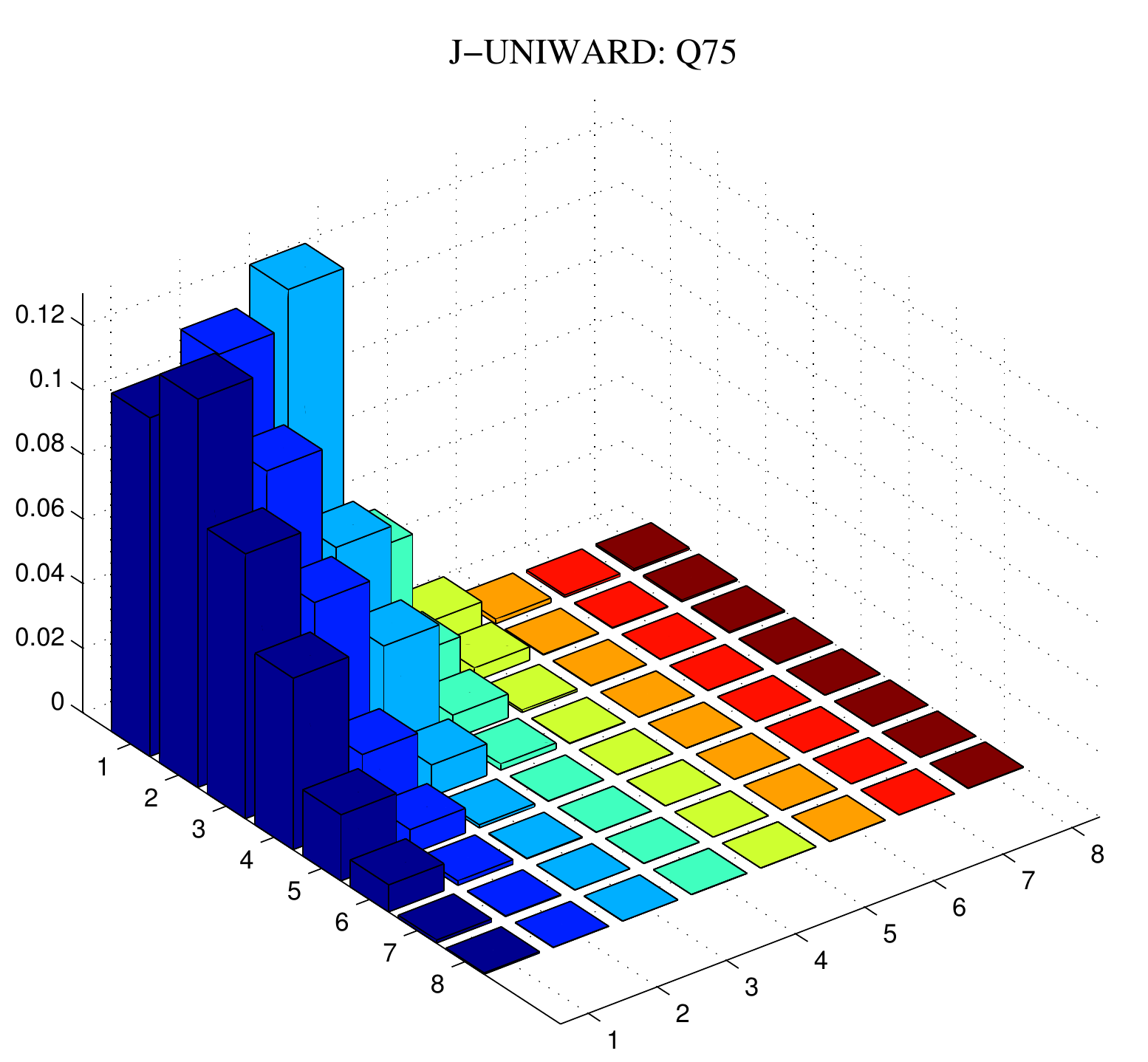}}
}
{
\subfigure[]{\label{fig:DCDT_Q75}\includegraphics[width=0.42\textwidth]{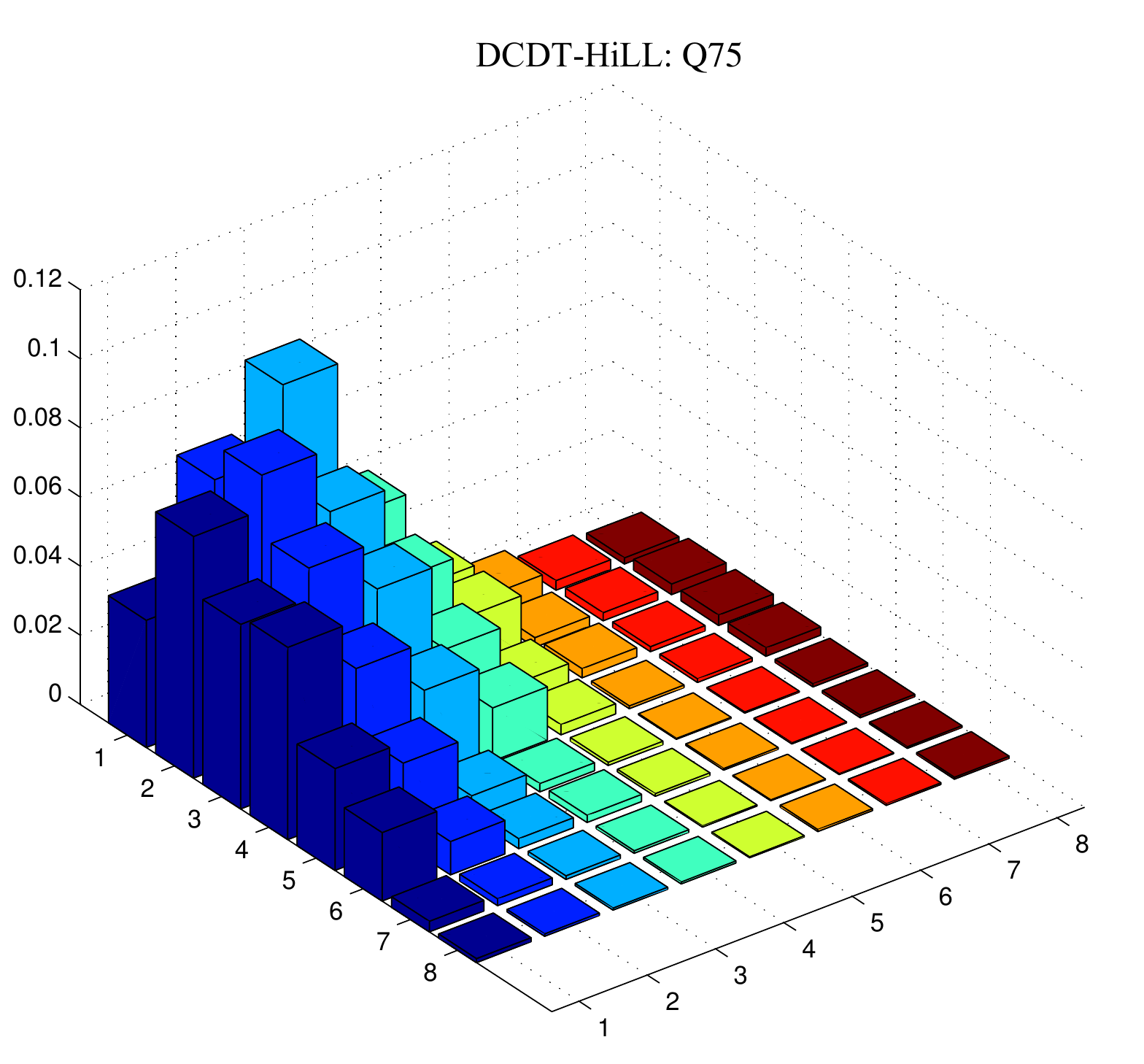}}
}
{
\subfigure[]{\label{fig:J-UNIWARD_Q95}\includegraphics[width=0.42\textwidth]{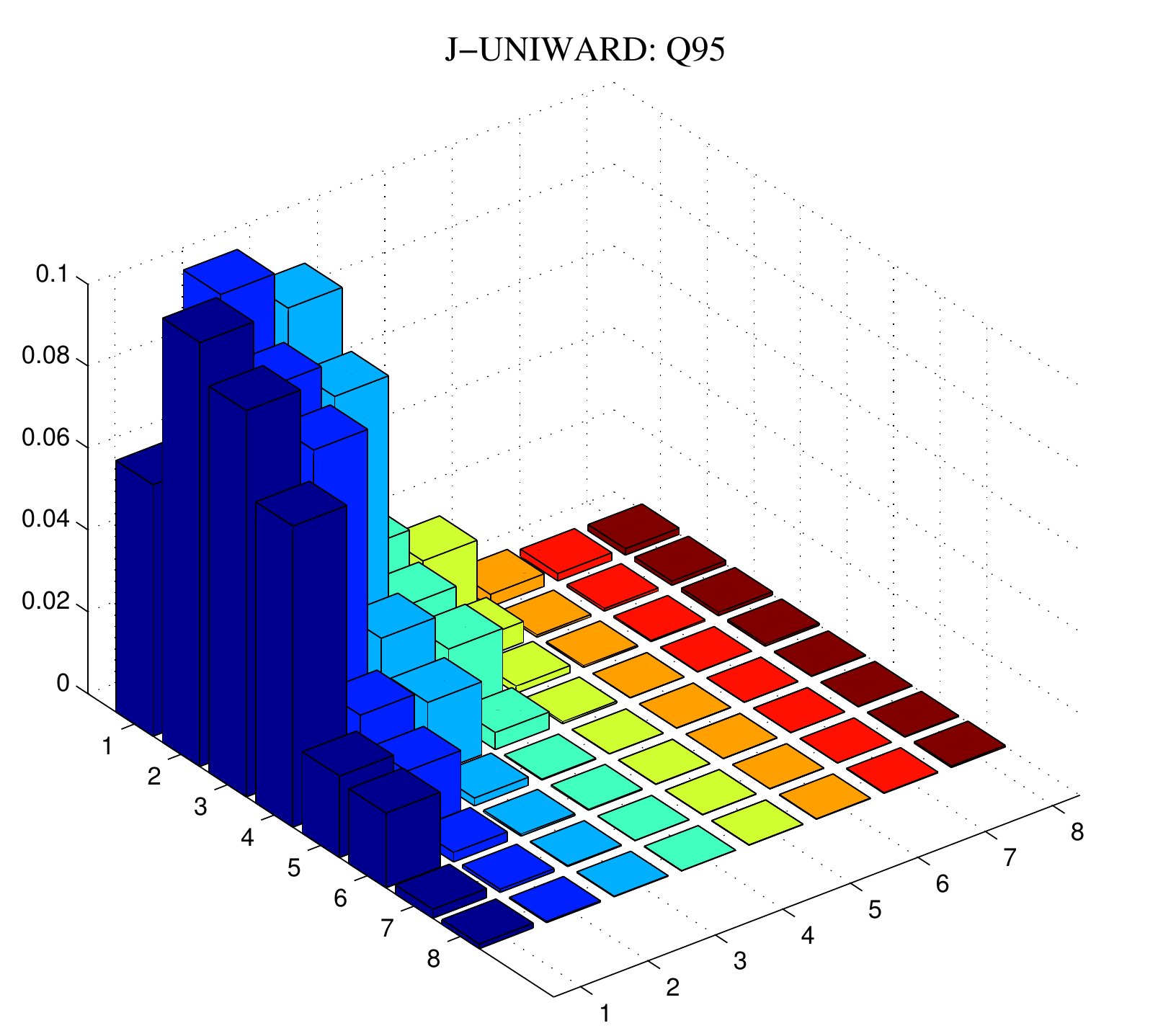}}
}
{
\subfigure[]{\label{fig:DCDT_Q95}\includegraphics[width=0.42\textwidth]{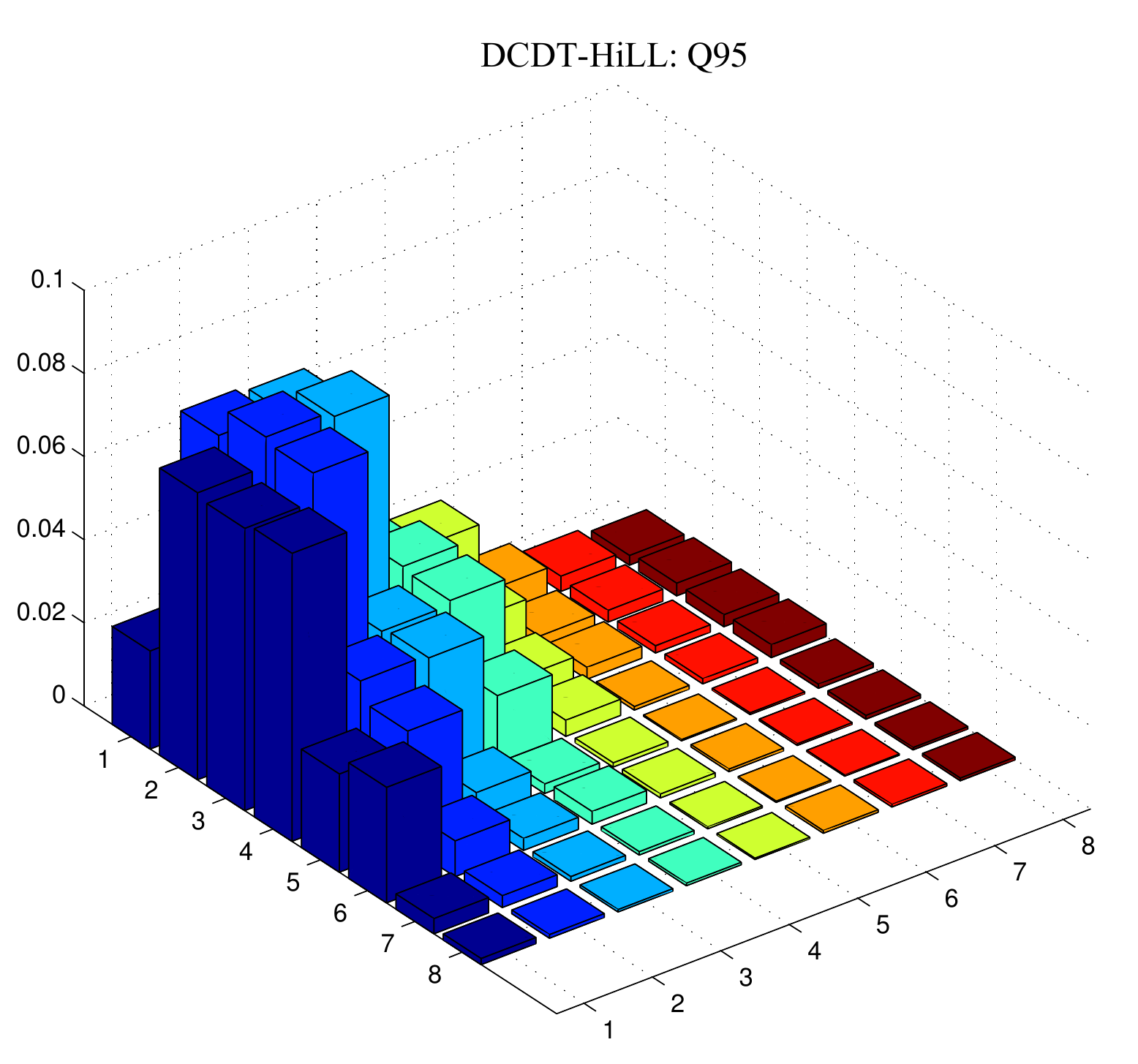}}
}
\caption{The statistical histogram for the percentage of average embedding modifications of 64 DCT modes under J-UNIWARD (Left) and DCDT-HiLL (Right) at relative payload 0.4 bpnzAC for Q75 (Top) and Q95 (Bottom).}\label{fig:mode_embedding_change}
\end{figure*}

It is noted that the existing content-adaptive spatial steganographic schemes, e.g., WOW \cite{WOW}, S-UNIWARD \cite{holub2014universal}, HiLL \cite{li2014a}, and MiPOD \cite{sedighi2016content}, are prone to embed messages in rich texture regions of the cover image. Therefore, the proposed DCDT may have a tendency to encourage more embedding modifications on mid-to-high frequency DCT coefficients, compared with the previous ones, e.g., J-UNIWARD. To validate this, we randomly select 2,000 covers images from BOSSBase ver1.01 \cite{bas2011break} at Q75 and Q95 separately and then perform embedding with J-UNIWARD and DCDT-HiLL{\footnote{Similar to the situation of BET-HiLL, it indicates that the DCDT scheme adopts HiLL as the spatial steganographic distortion cost function.} at relative payload 0.4 bpnzAC (bit per non-zero cover AC coefficient). As a result, the average embedding modification histograms of 64 DCT modes for the four stego sets are shown in Figure \ref{fig:mode_embedding_change}, indicating that whether on Q75 or Q95, DCDT-HiLL has more modifications on mid-to-high frequency DCT modes than J-UNIWARD. The distributions of embedding modifications on mid-to-high frequency DCT coefficients, however, should be well controlled, otherwise, the resulting spatial changes would become larger, especially at low QFs, which in turn make the embedding insecure. To tackle this issue, the distortion function in Eq. \eqref{eq:rho} is rewritten as the exponential form in Eq. \eqref{eq:rho_improve} below:
\begin{equation}\label{eq:rho_improve}
\rho _{a,b}^{m,n} = f\left( {{{\mathop{\bf{d}}\nolimits} _{m,n}},{{\mathop{\bf{s}}\nolimits}_{a,b}},p} \right) = \sum\limits_{i = 1}^8 {\sum\limits_{j = 1}^8 {{{\left({d_{m,n}{(i,j)}} \right)}^p} \cdot \left| {s_{a,b}{(i,j)}} \right|} },
\end{equation}
where $p$ is the exponent parameter, which is used to flexibly adjust the embedding distributions among different DCT blocks. With the distortion function defined in Eq. \eqref{eq:rho_improve}, the proposed JPEG steganographic scheme is developed under the STC-based minimal distortion embedding framework as shown in Fig. \ref{fig:flowchart}.
\begin{figure}[thbp]
\centering
{\includegraphics[width=0.8\textwidth]{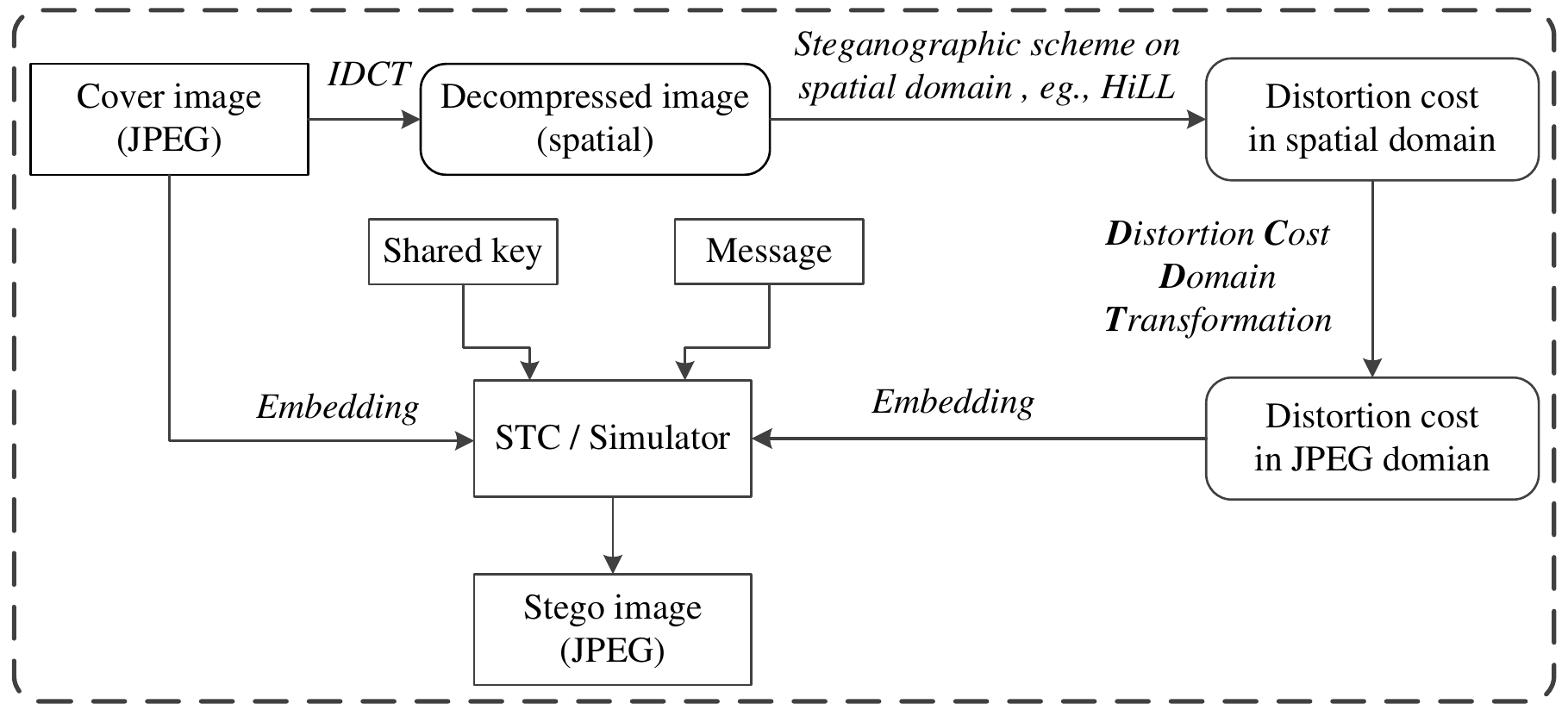}}
\caption{ The diagram of the proposed scheme. (IDCT for Inverse Discreet Cosine Transform)}\label{fig:flowchart}
\end{figure}

\subsection{Extension to mutually dependent embedding for the proposed scheme}
In practical applications, multiple DCT coefficients in one DCT block may be modified simultaneously, then the Eq. \eqref{eq:J2S} will be updated as:
\begin{equation}\label{eq:J2S_mutually}
{\bf{s}}{\rm{ = }}\left[ {{{\bf{A}}^T} * ({\bf{t}} \cdot {\bf{q}}) * {\bf{A}}} \right],
\end{equation}
where ${\bf{t}}$ represents the simultaneous modifications on multiple DCT modes in one ${8 \times 8}$ DCT block, and ${\bf{q}}$ is the corresponding quantization step matrix. As thus, the influence of embedding modifications in spatial domain would be mutually dependent. Recently, several mutually dependent embedding schemes have been developed, which are generally called the {\textbf{S}}ynchronizing {\textbf{M}}odification {\textbf{D}}irection (SMD) strategy, e.g., CMD \cite{CMD}, Synch \cite{Synch}, ASYMM \cite{ASYMM} and Dejoin \cite{Dejoin}\cite{Dejoin-J}, and by which, we can extend our proposed DCDT-HiLL to mutually dependent embedding, as illustrated in Figure \ref{fig:non_additive}. First of all, similar to the SMD embedding schemes, we perform the trial ternary embedding with DCDT-HiLL at the given payload, and then the embedding modification result, i.e., +1, 0, -1, for all the DCT coefficients in cover image can be accordingly obtained, for brevity, we call it the embedding modification map and denoted by $M$. Without loss of generality, we take the $k^{th}$ $8 \times 8$ block $m_k$ of $M$ in alignment with the DCT block of the cover for example, and record the indexes of the non-zero elements inside block $m_k$ as well as their number ($n_k$). Subsequently, we traverse all the non-zero elements inside $m_k$ for their adjustment of modification direction (+1/-1), which will then generates $2^{n_k}$ embedding modification candidate blocks. Likewise, for each of these candidates, the corresponding JPEG embedding distortion cost can be obtained by
\begin{equation}\label{eq:rho_improve_mutually}
\rho _{k} = \sum\limits_{i = 1}^8 {\sum\limits_{j = 1}^8 {{{\left({d_{k}{(i,j)}} \right)}^p} \cdot \left| {s_{k}{(i,j)}} \right|} },
\end{equation}
where $d_{k}{(i,j)}$ represents the spatial distortion cost of the ${(i,j)^{th}}$ pixel in the $k^{th}$ block of decompressed image for unit embedding change, $\left| {s_{k}{(i,j)}} \right|$ indicates the spatial mutually dependent embedding changes on the ${(i,j)^{th}}$ pixel in the corresponding block, and which can be obtained by Eq. \eqref{eq:J2S_mutually}.

After that, the optimal embedding modification block can be then obtained by finding out the one which yields the minimum embedding distortion cost among the $2^{n_k}$ candidate blocks, and in this way, the optimal embedding modification map $M'$ will be obtained after we traverse all the blocks in $M$. Finally, referring to $M'$, we appropriately update the original distortion cost $\rho$ calculated by DCDT-HiLL, and then use the updated distortion cost $\rho '$, which is referred to as DCDT-HiLL\_ud, to perform ternary embedding once again. Similar to the SMD strategy, the proposed distortion cost updating has the following definition:
\begin{equation}\label{eq:updating1}
{\rho '}_{i,j}^{+} = \left\{ \begin{array}{l}
{\rho _{i,j}}/v,M'_{i,j} = + 1\\
{\rho _{i,j}}*v,M'_{i,j} = - 1\\
{\rho _{i,j}},M'_{i,j} = 0
\end{array} \right.,
{\rho '}_{i,j}^{-} =\left\{ \begin{array}{l}
{\rho _{i,j}}/v,M'_{i,j} = - 1\\
{\rho _{i,j}}*v,M'_{i,j} = + 1\\
{\rho _{i,j}},M'_{i,j} = 0
\end{array} \right.,
\end{equation}
where the subscript $\{i,j\}$ stands for the index of DCT coefficient $x_{i,j}$, ${\rho '}_{i,j}^{+}$ and ${\rho '}_{i,j}^{-}$ are the updated distortion costs for modification $x_{i,j}+1$ and $x_{i,j}-1$, respectively, and $v$ is the penalty factor. The implementation of mutually dependent embedding does improve the performance at the cost of exponential complexity, therefore, unless otherwise specified, all the experiments in this paper are carried out with Mutually Independent (MI) embedding. And to the best of our knowledge, the MI embedding has also been used in J-UNIWARD and GUED with superior security performance.

\begin{figure*}[thbp]
\centering
{\psfig{figure=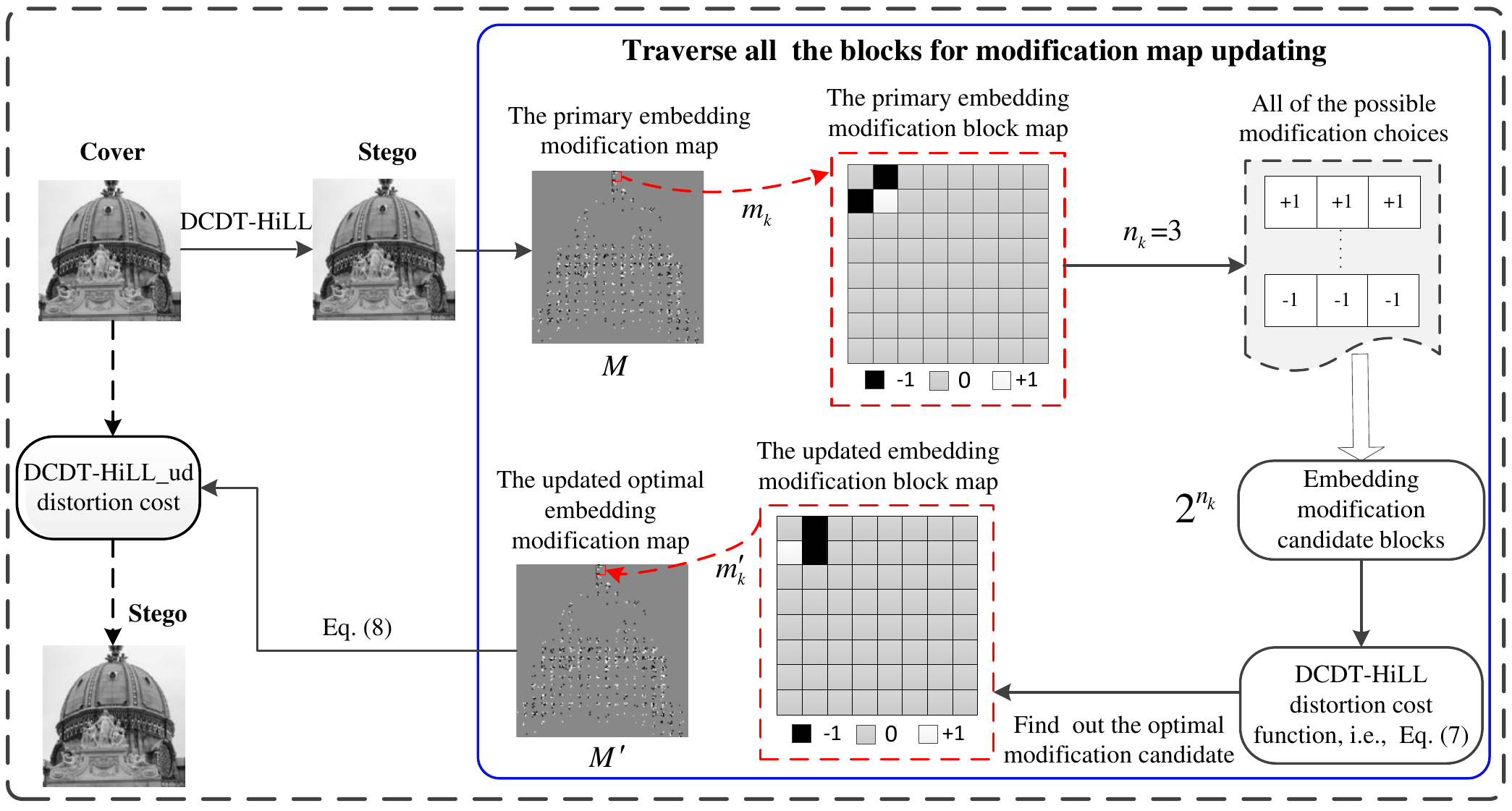,width=1\textwidth}}
\caption{The diagram of the mutually dependent embedding extension of the proposed scheme, including primary embedding, embedding modification map optimization, distortion cost updating and secondary embedding.}\label{fig:non_additive}
\end{figure*}

\section{Experimental results and analysis}
\subsection{Experiment setups}
All the experiments in this section are carried out on image database BOSSBase ver1.01 \cite{bas2011break} and BOWS2 \cite{BOWS2}, and both of them contain 10,000 gray-scale images of size $512 \times 512 \times 8$ bits. All the images in each database will be compressed by the JPEG Toolbox \cite{Jpeg_Toolbox} at different Quality Factors (QFs) to obtain various JPEG image sets, and for each JPEG image set, one half of them are used for training, while others for testing. To differentiate among various dataset, in the following, we use the syntax of names for JPEG image set following the convention: $name = {\{primary{\_}dataset\}\{J\}\{QF\}}$, where $primary{\_}dataset$ indicates the candidate image database, e.g., BOSSBase and BOWS2, $J$ stands for JPEG compression option, and $QF$ is the quality factor used in JPEG compression.

Several SOTA universal JPEG steganalyzers, including CC-JRM-22,510D \cite{kodovsky2012steganalysis}, GFR-17,000D \cite{song2015steganalysis} and its selection-channel-aware version SCA-GFR-17,000D \cite{denemark2016steganalysis}, are employed to evaluate the empirical security performance of the involved JPEG steganographic schemes, where the binary classifier is trained by the Fisher Linear Discriminants (FLD) ensemble \cite{kodovsky2012ensemble} with default settings. The classification error probability $P_E$ of FLD ensemble classifier, corresponding to the empirical security performance of the tested JPEG steganographic scheme, is reported by the mean value of the ensemble's testing errors based on ten times of randomly testing, and all the experiments are simulated at the corresponding payload distortion bound for relative payloads $\alpha$$\in$$\{0.1,0.2,0.3,0.4,0.5\}$ bpnzAC.

\subsection{Determining the optimal exponent parameter $p$ in DCDT-HiLL}
Since the exponent parameter $p$ in Eq. \eqref{eq:rho_improve} can be used to adjust the distributions of embedding modifications among DCT blocks, there should be an optimal $p$ setting for given steganalyzer, QF, and relative payload. To determine the $p$ in DCDT-HiLL for given QF and relative payload w.r.t. three SOTA JPEG steganalyzers CC-JRM, GFR, and SCA-GFR, we randomly select 5,000 images from BOSSBase with given QF, in which 2,500 JPEG images are used for training, while others for testing. We set $p$ in the range of [0.3,1.5] and search with interval 0.1 to find the optimal $p^*$ corresponding to the maximum classification error probability $P_E^*$ at given relative payload $\alpha$ for each of the three tested steganalyzers. The $p^*$ for GFR versus relative payloads on BOSSbaseJ75 and BOSSbaseJ95 are illustrated in Fig. \ref{fig:p_GFR}, it shows that the optimal parameters $p$ is nearly irrelevant to relative payloads, and for simplicity, we finally set $p^*$ as 0.7 and 1.1 for Q75 and Q95, respectively. Similarly, the optimal parameters $p^*$ for Q75 and Q95 w.r.t. other tested steganalyzers can be obtained as well, which are all summarized in Table \ref{tab:optimal_p}.
\begin{figure*} [h]
\centering
{
\subfigure[]{\label{fig:p_GFR_Q75}\includegraphics[width=0.49\textwidth]{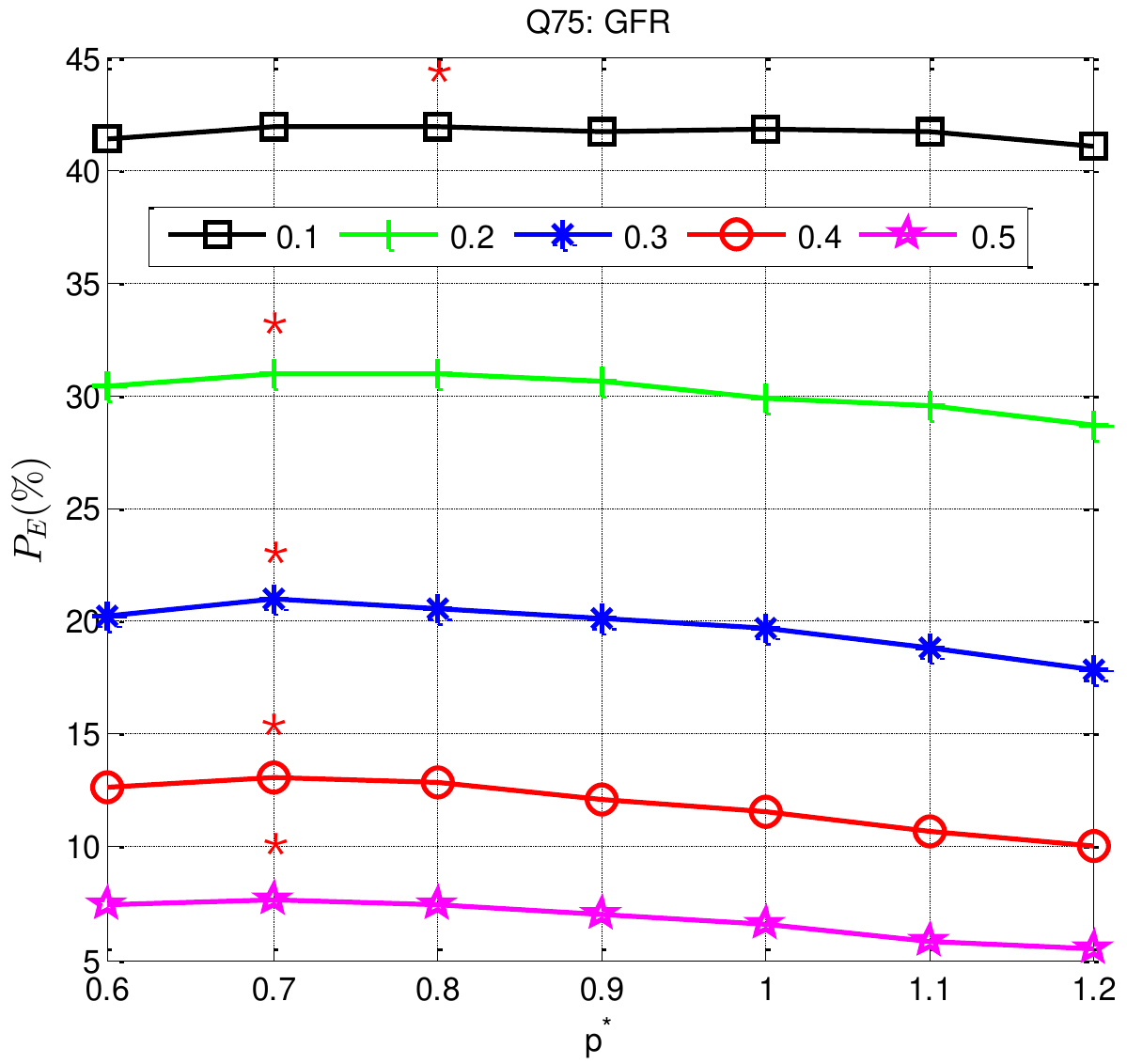}}
}
{
\subfigure[]{\label{fig:p_GFR_Q95}\includegraphics[width=0.49\textwidth]{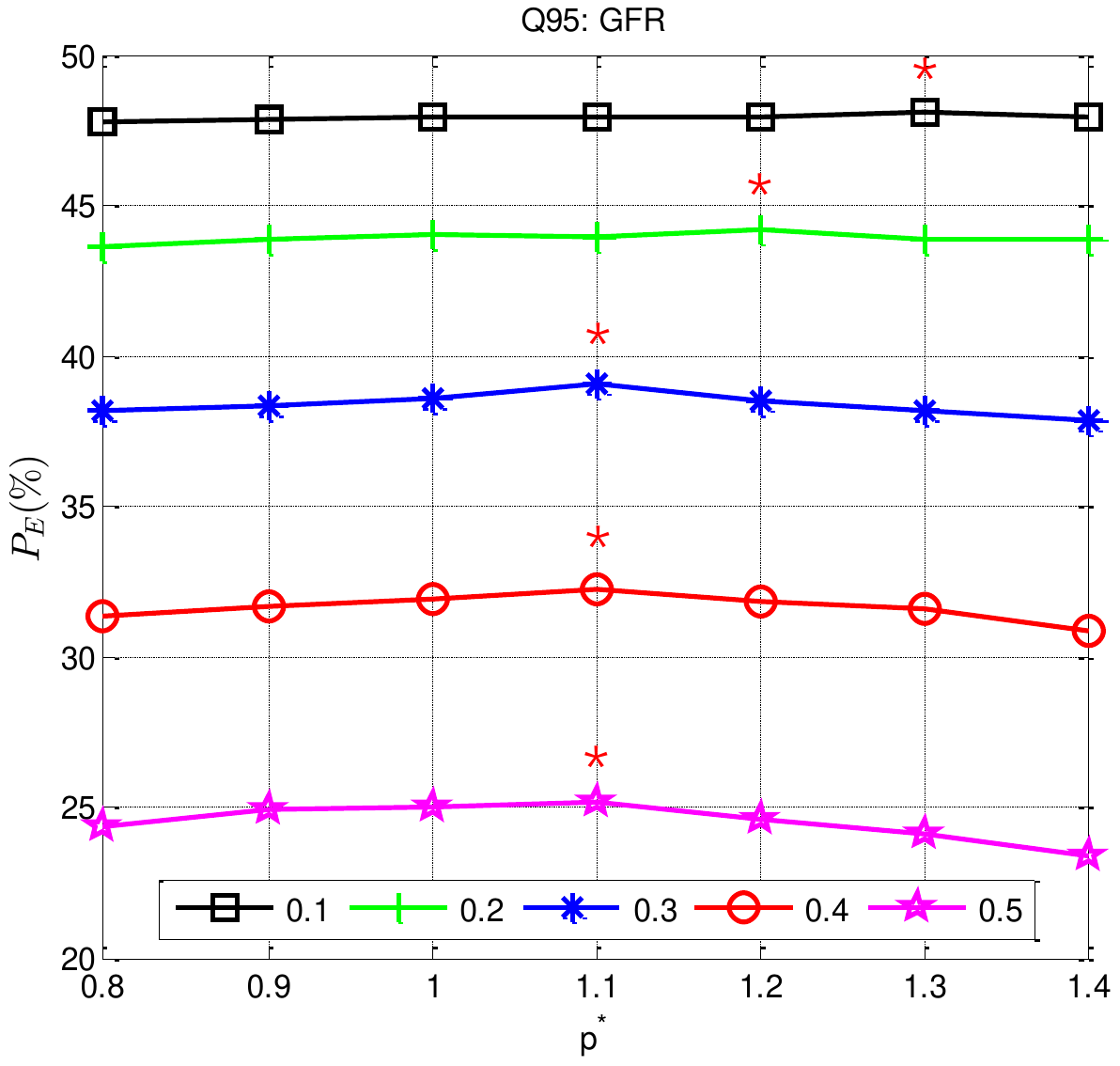}}
}
\caption{(a) and (b) are the classification error probability $P_E$ associated with different $p$ towards steganalyzer GFR versus relative payloads $\alpha$ on BOSSbaseJ75 and BOSSbaseJ95, respectively. The optimal results are indicated with symbol $*$ in the figure. }\label{fig:p_GFR}
\end{figure*}

\begin{table}[h]
\renewcommand\arraystretch{1}
\centering
\caption{The optimal parameter $p$ in DCDT-HiLL for all the involved steganalyzers CC-JRM, GFR and SCA-GFR on BOSSBaseJ75 and BOSSBaseJ95.}\label{tab:optimal_p}
\begin{tabular}{cccc}
  \toprule[1pt]
    \multirow{2}{*}{QF} & \multicolumn{3}{c}{Steganalyzer}\\
     \cmidrule(lr){2-4}
     \centering
     &CC-JRM &GFR &SCA-GFR\\
    \midrule[0.8pt]
        75 	&$0.7$ &$0.7$ &$0.5$\\
    \midrule[0.8pt]
        95 	&$0.9$ &$1.1$ &$0.9$\\
    \bottomrule[1pt]
\end{tabular}
\end{table}
\subsection{The performance of the proposed JPEG steganographic scheme with various spatial steganographic distortion cost functions}
To verify the effectiveness of the proposed JPEG steganographic scheme equipped with spatial steganographic distortion cost function HiLL, another two SOTA spatial distortion cost functions S-UNIWARD and MiPOD are then used for comparison. Similar to the procedure of HiLL in determining the optimal $p$, we can easily obtain the optimal parameters $p$ for S-UNIWARD and MiPOD on BOSSBaseJ75 and BOSSBaseJ95 w.r.t. CC-JRM and GFR, which are shown in Table \ref{tab:optimal_p_2}. Then, we compare the security performance of DCDT-HiLL with DCDT-S-UNIWARD and DCDT-MiPOD against the detection of CC-JRM and GFR on BOSSBaseJ75 and BOSSBaseJ95 at their corresponding optimal $p$ as shown in Table \ref{tab:Security performance_different_spatial_scheme}. It is observed that the proposed DCDT equipped with HiLL exhibits the best security performance, and it is adopted in the rest of the paper unless otherwise specified.

\begin{table}[h]
\renewcommand\arraystretch{1}
\centering
\caption{The optimal parameter $p$ in DCDT-S-UNIWARD and DCDT-MiPOD for steganalyzers CC-JRM and GFR on BOSSBaseJ75 and BOSSBaseJ95.}\label{tab:optimal_p_2}
\begin{tabular}{cccccc}
  \toprule[1pt]
    \multirow{2}{*}{QF} &\multicolumn{2}{c}{S-UNIWARD} & &\multicolumn{2}{c}{MiPOD}\\
     \cmidrule(lr){2-6}
     \centering
     &CC-JRM &GFR  & &CC-JRM &GFR\\
    \midrule[0.8pt]
        75 	&$1.2$ &$1.3$  &  &$0.7$ &$0.7$ \\
    \midrule[0.8pt]
        95 	&$1.9$ &$2.0$  &  &$0.7$ &$0.7$ \\
    \bottomrule[1pt]
\end{tabular}
\end{table}

\begin{table}[thbp]
\renewcommand\arraystretch{1}
\caption{Classification error probability ${P_E}$ (in \%) of DCDT-S-UNIWARD, DCDT-MiPOD and DCDT-HiLL for CC-JRM and GFR versus relative payloads on BOSSBaseJ75 and BOSSBaseJ95.}\label{tab:Security performance_different_spatial_scheme}
\centering
\begin{tabular}{cccccccc}
\toprule [1pt]
\centering
 \multirow{2}{*}{Steganalyzer}&\multirow{2}{*}{QF}&\multirow{2}{*}{Scheme}& \multicolumn{5}{c}{Relative payload $\alpha$ (bpnzAC)}\\
  \cmidrule(lr){4-8}
 \centering
  & & &0.1 &0.2 &0.3 &0.4 &0.5\\
  \midrule[0.6pt]
  \centering
  \multirow{6}{*}{CC-JRM} &\multirow{3}{*}{75} &DCDT-S-UNIWARD  &${46.48}$ &${39.66}$ &${32.01}$ &${23.68}$ &${16.52}$\\
                           & &DCDT-MiPOD   &${45.97}$ &${39.27}$ &${31.27}$ &${23.20}$ &${15.86}$\\
                           & &DCDT-HiLL   &$\bf{46.62}$	&$\bf{40.10}$ &$\bf{32.25}$ &$\bf{24.10}$ &$\bf{17.02}$\\
                      \cmidrule[0.6pt]{2-8}
                      &\multirow{3}{*}{95}     &DCDT-S-UNIWARD   &${49.39}$ &${47.39}$ &${43.77}$ &${38.49}$ &${31.19}$\\
                         &  &DCDT-MiPOD   &${49.27}$ &${46.99}$ &${42.60}$ &${35.96}$ &${28.52}$\\
                         &  &DCDT-HiLL    &${\bf49.44}$ &$\bf{47.84}$ &$\bf{45.05}$ &$\bf{40.12}$ &$\bf{33.53}$ \\
  \midrule[0.6pt]
  \centering
  \multirow{6}{*}{GFR} &\multirow{3}{*}{75} &DCDT-S-UNIWARD   &${41.08}$ &${29.47}$ &${18.89}$ &${11.12}$ &${6.07}$\\
                           & &DCDT-MiPOD   &${39.96}$ &${27.58}$ &${17.20}$ &${9.90}$ &${5.33}$\\
                           & &DCDT-HiLL   &$\bf{41.30}$	&$\bf{29.95}$ &$\bf{19.61}$ &$\bf{12.06}$ &$\bf{6.80}$ \\
                      \cmidrule[0.6pt]{2-8}
                      &\multirow{3}{*}{95}     &DCDT-S-UNIWARD   &${47.69}$ &${43.06}$ &${37.01}$ &${29.35}$ &${21.67}$\\
                         &  &DCDT-MiPOD   &${47.50}$ &${42.62}$ &${36.18}$ &${28.56}$ &${20.71}$\\
                         &  &DCDT-HiLL   &$\bf{47.94}$ &$\bf{43.67}$ &$\bf{38.16}$ &$\bf{31.22}$ &$\bf{23.80}$\\
\toprule[1pt]
\end{tabular}
\end{table}

\subsection{Performance comparison of the proposed DCDT-HiLL with other SOTA JPEG steganographic schemes}
We then compare the security performance of the proposed DCDT-HiLL with other SOTA JPEG steganographic schemes, e.g., UERD, J-UNIWARD, GUED, and BET-HiLL at different relative payloads on BOSSBaseJ75 and BOSSBaseJ95, which are summarized in Table \ref{tab:Security performance_ori_Q75} and \ref{tab:Security performance_ori_Q95}, respectively. For brevity, the results of our proposed DCDT-HiLL (except \emph{DCDT-HiLL-pro}) in Table \ref{tab:Security performance_ori_Q75} and \ref{tab:Security performance_ori_Q95} are obtained with the optimal parameter setting for SCA-GFR (i.e., $p$=0.5 and $p$=0.9 for Q75 and Q95, respectively.). This is because SCA-GFR is the most effective steganalyzer and the performance of the proposed DCDT-HiLL with the same parameter setting won't change much as justified by our experiments.

As shown in Table \ref{tab:Security performance_ori_Q75} and \ref{tab:Security performance_ori_Q95}, it is observed that for steganalyzer SCA-GFR, the proposed DCDT-HiLL achieves an overall superior performance than UERD, J-UNIWARD, and GUED. In addition, DCDT-HiLL also consistently outperforms BET-HiLL by a clear margin (increase the $P_E$ by 1.4\%-2.1\% on average) for JPEG images of Q75, and shows comparable performance with BET-HiLL for Q95.

For the steganalyzer GFR, however, although our proposed DCDT-HiLL still exhibits excellent performance compared with other competing schemes except BET-HiLL for JPEG images of Q95, the performance gains are significantly narrowed for Q75. And it only shows comparable or slightly inferior performance than BET-HiLL whether for JPEG images of Q75 or Q95. The following two reasons may contribute to the performance degradation. One is that the suboptimal parameter setting for GFR. When the optimal parameter setting for GFR under Q75 is adopted, i.e., $p$=0.7 (\emph{DCDT-HiLL-pro}), the performance of the proposed DCDT-HiLL is indeed improved as illustrated in Table \ref{tab:Security performance_ori_Q75}. The other is the assumption of mutually independent embedding, which will be discussed later. Note that the effect of the quantization step, the embedding influence in the spatial domain for Q75 is much greater than that of Q95, which may lead to the performance decline for Q75 compared with the one for Q95.

\begin{table}[thbp]
\renewcommand\arraystretch{1}
\caption{Classification error probability ${P_E}$ (in \%) of the involved JPEG steganographic schemes for CC-JRM, GFR and SCA-GFR versus relative payloads on BOSSBaseJ75.} \label{tab:Security performance_ori_Q75}
\centering
\begin{threeparttable}
\begin{tabular}{ccccccc}
\toprule [1pt]
 \multirow{2}{*}{Steganalyzer}&\multirow{2}{*}{Scheme}& \multicolumn{5}{c}{Relative payload $\alpha$ (bpnzAC)}\\
  \cmidrule(lr){3-7}
 \centering
  & &0.1&0.2&0.3&0.4&0.5\\
  \midrule[0.6pt]
  \centering

   \multirow{6}{*}{{CC-JRM}$^{3rd}$}  &\multicolumn{1}{>{\columncolor{1mygray6}}l}{UERD}      &\multicolumn{1}{>{\columncolor{1mygray6}}l}{$45.89$}	&\multicolumn{1}{>{\columncolor{1mygray6}}l}{$38.93$} &\multicolumn{1}{>{\columncolor{1mygray6}}l}{$30.91$} &\multicolumn{1}{>{\columncolor{1mygray6}}l}{$23.22$} &\multicolumn{1}{>{\columncolor{1mygray6}}l}{$16.53$} \\
                            &\multicolumn{1}{>{\columncolor{1mygray2}}l}{J-UNIWARD} &\multicolumn{1}{>{\columncolor{1mygray2}}l}{$47.10$}	&\multicolumn{1}{>{\columncolor{1mygray2}}l}{$41.25$} &\multicolumn{1}{>{\columncolor{1mygray2}}l}{$34.00$} &\multicolumn{1}{>{\columncolor{1mygray2}}l}{$26.83$} &\multicolumn{1}{>{\columncolor{1mygray2}}l}{$19.29$} \\
                            &\multicolumn{1}{>{\columncolor{1mygray1}}l}{GUED} &\multicolumn{1}{>{\columncolor{1mygray1}}l}{${47.27}$}	&\multicolumn{1}{>{\columncolor{1mygray1}}l}{${41.33}$} &\multicolumn{1}{>{\columncolor{1mygray1}}l}{${34.83}$} &\multicolumn{1}{>{\columncolor{1mygray1}}l}{${27.18}$} &\multicolumn{1}{>{\columncolor{1mygray1}}l}{${20.08}$} \\
                            &\multicolumn{1}{>{\columncolor{1mygray3}}l}{BET-HiLL}  &\multicolumn{1}{>{\columncolor{1mygray3}}l}{$46.76$}	&\multicolumn{1}{>{\columncolor{1mygray3}}l}{$40.57$} &\multicolumn{1}{>{\columncolor{1mygray3}}l}{$32.71$} &\multicolumn{1}{>{\columncolor{1mygray3}}l}{$24.74$} &\multicolumn{1}{>{\columncolor{1mygray3}}l}{$17.36$} \\
                            &\multicolumn{1}{>{\columncolor{1mygray5}}l}{DCDT-HiLL}     &\multicolumn{1}{>{\columncolor{1mygray5}}l}{$46.14$}	&\multicolumn{1}{>{\columncolor{1mygray5}}l}{$39.53$} &\multicolumn{1}{>{\columncolor{1mygray5}}l}{$31.44$} &\multicolumn{1}{>{\columncolor{1mygray5}}l}{$23.39$} &\multicolumn{1}{>{\columncolor{1mygray5}}l}{$16.62$} \\
                            &\multicolumn{1}{>{\columncolor{1mygray4}}l}{\emph{DCDT-HiLL-pro}}      &\multicolumn{1}{>{\columncolor{1mygray4}}l}{$\emph{46.62}$}	&\multicolumn{1}{>{\columncolor{1mygray4}}l}{$\emph{40.10}$} &\multicolumn{1}{>{\columncolor{1mygray4}}l}{$\emph{32.25}$} &\multicolumn{1}{>{\columncolor{1mygray4}}l}{$\emph{24.10}$} &\multicolumn{1}{>{\columncolor{1mygray4}}l}{$\emph{17.02}$} \\

  \midrule[0.6pt]
  \centering
   \multirow{6}{*}{GFR$^{2nd}$}     &\multicolumn{1}{>{\columncolor{2mygray6}}l}{UERD}      &\multicolumn{1}{>{\columncolor{2mygray6}}l}{$39.97$}	&\multicolumn{1}{>{\columncolor{2mygray6}}l}{$27.80$} &\multicolumn{1}{>{\columncolor{2mygray6}}l}{$18.01$} &\multicolumn{1}{>{\columncolor{2mygray5}}l}{$10.47$} &\multicolumn{1}{>{\columncolor{2mygray5}}l}{$6.05$} \\
                            &\multicolumn{1}{>{\columncolor{2mygray5}}l}{J-UNIWARD} &\multicolumn{1}{>{\columncolor{2mygray5}}l}{$41.38$}	&\multicolumn{1}{>{\columncolor{2mygray5}}l}{$28.96$} &\multicolumn{1}{>{\columncolor{2mygray5}}l}{$18.29$} &\multicolumn{1}{>{\columncolor{2mygray5}}l}{$10.46$} &\multicolumn{1}{>{\columncolor{2mygray6}}l}{$5.58$} \\
                            &\multicolumn{1}{>{\columncolor{2mygray3}}l}{GUED}      &\multicolumn{1}{>{\columncolor{2mygray3}}l}{$41.57$}	&\multicolumn{1}{>{\columncolor{2mygray3}}l}{$29.93$} &\multicolumn{1}{>{\columncolor{2mygray3}}l}{$19.13$} &\multicolumn{1}{>{\columncolor{2mygray3}}l}{$11.14$} &\multicolumn{1}{>{\columncolor{2mygray3}}l}{$6.10$} \\
                            &\multicolumn{1}{>{\columncolor{2mygray1}}l}{BET-HiLL}  &\multicolumn{1}{>{\columncolor{2mygray1}}l}{${41.95}$}	&\multicolumn{1}{>{\columncolor{2mygray1}}l}{${31.32}$} &\multicolumn{1}{>{\columncolor{2mygray1}}l}{${21.18}$} &\multicolumn{1}{>{\columncolor{2mygray1}}l}{${13.38}$} &\multicolumn{1}{>{\columncolor{2mygray1}}l}{${7.56}$} \\
                            &\multicolumn{1}{>{\columncolor{2mygray4}}l}{DCDT-HiLL}     &\multicolumn{1}{>{\columncolor{2mygray4}}l}{$40.85$}	&\multicolumn{1}{>{\columncolor{2mygray4}}l}{$29.33$} &\multicolumn{1}{>{\columncolor{2mygray4}}l}{$18.62$} &\multicolumn{1}{>{\columncolor{2mygray4}}l}{$10.97$} &\multicolumn{1}{>{\columncolor{2mygray4}}l}{$6.27$} \\
                             &\multicolumn{1}{>{\columncolor{2mygray2}}l}{\emph{DCDT-HiLL-pro}}     &\multicolumn{1}{>{\columncolor{2mygray2}}l}{$\emph{41.30}$}	&\multicolumn{1}{>{\columncolor{2mygray2}}l}{$\emph{29.95}$} &\multicolumn{1}{>{\columncolor{2mygray2}}l}{$\emph{19.61}$} &\multicolumn{1}{>{\columncolor{2mygray2}}l}{$\emph{12.06}$} &\multicolumn{1}{>{\columncolor{2mygray2}}l}{$\emph{6.80}$} \\

  \midrule[0.6pt]
  \centering
   \multirow{5}{*}{SCA-GFR$^{1st}$} &\multicolumn{1}{>{\columncolor{3mygray5}}l}{UERD}      &\multicolumn{1}{>{\columncolor{3mygray5}}l}{$32.14$}	&\multicolumn{1}{>{\columncolor{3mygray5}}l}{$21.03$} &\multicolumn{1}{>{\columncolor{3mygray4}}l}{$13.64$} &\multicolumn{1}{>{\columncolor{3mygray2}}l}{$8.57$} &\multicolumn{1}{>{\columncolor{3mygray2}}l}{$5.04$} \\
                            &\multicolumn{1}{>{\columncolor{3mygray2}}l}{J-UNIWARD} &\multicolumn{1}{>{\columncolor{3mygray3}}l}{$35.98$}	&\multicolumn{1}{>{\columncolor{3mygray3}}l}{$23.35$} &\multicolumn{1}{>{\columncolor{3mygray2}}l}{$14.15$} &\multicolumn{1}{>{\columncolor{3mygray4}}l}{$8.03$} &\multicolumn{1}{>{\columncolor{3mygray3}}l}{$4.47$} \\
                            &\multicolumn{1}{>{\columncolor{3mygray2}}l}{GUED}      &\multicolumn{1}{>{\columncolor{3mygray2}}l}{$36.55$}	&\multicolumn{1}{>{\columncolor{3mygray3}}l}{$23.20$} &\multicolumn{1}{>{\columncolor{3mygray5}}l}{$13.59$} &\multicolumn{1}{>{\columncolor{3mygray5}}l}{$7.85$} &\multicolumn{1}{>{\columncolor{3mygray4}}l}{$4.42$} \\
                            &\multicolumn{1}{>{\columncolor{3mygray4}}l}{BET-HiLL}  &\multicolumn{1}{>{\columncolor{3mygray4}}l}{$34.71$}	&\multicolumn{1}{>{\columncolor{3mygray4}}l}{$22.55$} &\multicolumn{1}{>{\columncolor{3mygray3}}l}{$13.98$} &\multicolumn{1}{>{\columncolor{3mygray4}}l}{$8.06$} &\multicolumn{1}{>{\columncolor{3mygray5}}l}{$4.28$} \\
                            &\multicolumn{1}{>{\columncolor{3mygray1}}l}{DCDT-HiLL}     &\multicolumn{1}{>{\columncolor{3mygray1}}l}{${36.85}$}	&\multicolumn{1}{>{\columncolor{3mygray1}}l}{${24.51}$} &\multicolumn{1}{>{\columncolor{3mygray1}}l}{${15.51}$} &\multicolumn{1}{>{\columncolor{3mygray1}}l}{${9.49}$} &\multicolumn{1}{>{\columncolor{3mygray1}}l}{${5.95}$} \\
\toprule[1pt]
\end{tabular}
\begin{tablenotes}
    \footnotesize
    \item $\dag$ {{the detectability of the steganalyzers in Table \ref{tab:Security performance_ori_Q75} and \ref{tab:Security performance_ori_Q95} follows: SCA-GFR $>$  GFR $ \gg $ CC-JRM}}.
    \item $\bot$ {{The darkness of the background in the Table \ref{tab:Security performance_ori_Q75} and \ref{tab:Security performance_ori_Q95} indicates the security of steganographic schemes, i.e.,  the darker the background, the higher the security of steganographic scheme}}.
\end{tablenotes}
\end{threeparttable}
\end{table}

\begin{table}[thbp]
\renewcommand\arraystretch{1}
\caption{Classification error probability ${P_E}$ (in \%) of the involved JPEG steganographic schemes for CC-JRM, GFR and SCA-GFR versus relative payloads on BOSSBaseJ95.}\label{tab:Security performance_ori_Q95}
\centering
\begin{tabular}{ccccccc}
\toprule [1pt]
 \multirow{2}{*}{Steganalyzer}&\multirow{2}{*}{Scheme}& \multicolumn{5}{c}{Relative payload $\alpha$ (bpnzAC)}\\
 \cmidrule(lr){3-7}
 \centering
  & &0.1&0.2&0.3&0.4&0.5\\
  \midrule[0.6pt]
  \centering

   \multirow{5}{*}{CC-JRM$^{3rd}$}  &\multicolumn{1}{>{\columncolor{4mygray5}}l}{UERD}      &\multicolumn{1}{>{\columncolor{4mygray5}}l}{$49.04$}	  &\multicolumn{1}{>{\columncolor{4mygray5}}l}{$46.57$}	&\multicolumn{1}{>{\columncolor{4mygray5}}l}{$42.02$}	&\multicolumn{1}{>{\columncolor{4mygray5}}l}{$35.97$}	&\multicolumn{1}{>{\columncolor{4mygray5}}l}{$29.04$} \\
                            &\multicolumn{1}{>{\columncolor{4mygray2}}l}{J-UNIWARD} &\multicolumn{1}{>{\columncolor{4mygray2}}l}{$49.55$}	  &\multicolumn{1}{>{\columncolor{4mygray2}}l}{$47.94$}	&\multicolumn{1}{>{\columncolor{4mygray2}}l}{$45.07$}	&\multicolumn{1}{>{\columncolor{4mygray2}}l}{$40.76$}	&\multicolumn{1}{>{\columncolor{4mygray2}}l}{$35.13$} \\
                            &\multicolumn{1}{>{\columncolor{4mygray1}}l}{GUED}      &\multicolumn{1}{>{\columncolor{4mygray1}}l}{${49.57}$}	  &\multicolumn{1}{>{\columncolor{4mygray1}}l}{${48.13}$} &\multicolumn{1}{>{\columncolor{4mygray1}}l}{${45.63}$} &\multicolumn{1}{>{\columncolor{4mygray1}}l}{${42.15}$} &\multicolumn{1}{>{\columncolor{4mygray1}}l}{${37.16}$} \\
                            &\multicolumn{1}{>{\columncolor{4mygray3}}l}{BET-HiLL}  &\multicolumn{1}{>{\columncolor{4mygray3}}l}{$49.51$}	&\multicolumn{1}{>{\columncolor{4mygray4}}l}{$47.74$}	&\multicolumn{1}{>{\columncolor{4mygray4}}l}{$44.92$}	&\multicolumn{1}{>{\columncolor{4mygray3}}l}{$40.56$}	&\multicolumn{1}{>{\columncolor{4mygray3}}l}{$34.30$} \\
                            &\multicolumn{1}{>{\columncolor{4mygray4}}l}{DCDT-HiLL}     &\multicolumn{1}{>{\columncolor{4mygray4}}l}{$49.44$}	&\multicolumn{1}{>{\columncolor{4mygray3}}l}{$47.84$}	&\multicolumn{1}{>{\columncolor{4mygray3}}l}{$45.05$}	&\multicolumn{1}{>{\columncolor{4mygray4}}l}{$40.12$}	&\multicolumn{1}{>{\columncolor{4mygray4}}l}{$33.53$} \\

  \midrule[0.6pt]
  \centering
   \multirow{5}{*}{GFR$^{2nd}$}     &\multicolumn{1}{>{\columncolor{5mygray5}}l}{UERD}      &\multicolumn{1}{>{\columncolor{5mygray5}}l}{$46.07$}	  &\multicolumn{1}{>{\columncolor{5mygray5}}l}{$39.62$}	&\multicolumn{1}{>{\columncolor{5mygray5}}l}{$32.45$}	&\multicolumn{1}{>{\columncolor{5mygray5}}l}{$24.68$}	&\multicolumn{1}{>{\columncolor{5mygray5}}l}{$17.84$} \\
                            &\multicolumn{1}{>{\columncolor{5mygray4}}l}{J-UNIWARD} &\multicolumn{1}{>{\columncolor{5mygray4}}l}{$47.55$}  &\multicolumn{1}{>{\columncolor{5mygray4}}l}{$42.74$}	&\multicolumn{1}{>{\columncolor{5mygray4}}l}{$35.88$}	&\multicolumn{1}{>{\columncolor{5mygray4}}l}{$28.17$}	&\multicolumn{1}{>{\columncolor{5mygray4}}l}{$20.54$} \\
                            &\multicolumn{1}{>{\columncolor{5mygray3}}l}{GUED}      &\multicolumn{1}{>{\columncolor{5mygray3}}l}{$47.24$}	  &\multicolumn{1}{>{\columncolor{5mygray3}}l}{$42.78$}  &\multicolumn{1}{>{\columncolor{5mygray3}}l}{$36.23$}    &\multicolumn{1}{>{\columncolor{5mygray3}}l}{$29.42$}     &\multicolumn{1}{>{\columncolor{5mygray3}}l}{$23.21$} \\
                            &\multicolumn{1}{>{\columncolor{5mygray1}}l}{BET-HiLL}  &\multicolumn{1}{>{\columncolor{5mygray1}}l}{${48.01}$}	&\multicolumn{1}{>{\columncolor{5mygray1}}l}{${43.91}$}	&\multicolumn{1}{>{\columncolor{5mygray1}}l}{${38.51}$}	&\multicolumn{1}{>{\columncolor{5mygray1}}l}{${31.82}$}	&\multicolumn{1}{>{\columncolor{5mygray1}}l}{${25.29}$} \\
                            &\multicolumn{1}{>{\columncolor{5mygray2}}l}{DCDT-HiLL}        &\multicolumn{1}{>{\columncolor{5mygray2}}l}{$47.57$}	&\multicolumn{1}{>{\columncolor{5mygray2}}l}{$43.40$}    &\multicolumn{1}{>{\columncolor{5mygray2}}l}{$37.90$}    &\multicolumn{1}{>{\columncolor{5mygray2}}l}{$31.35$}	&\multicolumn{1}{>{\columncolor{5mygray3}}l}{$23.84$} \\

  \midrule[0.6pt]
  \centering
   \multirow{5}{*}{SCA-GFR$^{1st}$} &\multicolumn{1}{>{\columncolor{6mygray7}}l}{UERD}      &\multicolumn{1}{>{\columncolor{6mygray7}}l}{$44.03$}	&\multicolumn{1}{>{\columncolor{6mygray7}}l}{$37.91$}	&\multicolumn{1}{>{\columncolor{6mygray6}}l}{$31.45$}	&\multicolumn{1}{>{\columncolor{6mygray6}}l}{$25.47$}	&\multicolumn{1}{>{\columncolor{6mygray7}}l}{$19.32$} \\
                            &\multicolumn{1}{>{\columncolor{6mygray3}}l}{J-UNIWARD} &\multicolumn{1}{>{\columncolor{6mygray3}}l}{$46.17$}	&\multicolumn{1}{>{\columncolor{6mygray3}}l}{$40.49$}	&\multicolumn{1}{>{\columncolor{6mygray4}}l}{$33.77$}	&\multicolumn{1}{>{\columncolor{6mygray4}}l}{$26.63$}	&\multicolumn{1}{>{\columncolor{6mygray5}}l}{$20.38$} \\
                            &\multicolumn{1}{>{\columncolor{6mygray4}}l}{GUED}      &\multicolumn{1}{>{\columncolor{6mygray6}}l}{$44.97$}	&\multicolumn{1}{>{\columncolor{6mygray6}}l}{$37.94$}    &\multicolumn{1}{>{\columncolor{6mygray7}}l}{$30.98$}    &\multicolumn{1}{>{\columncolor{6mygray7}}l}{$24.99$}    &\multicolumn{1}{>{\columncolor{6mygray6}}l}{$19.62$} \\
                            &\multicolumn{1}{>{\columncolor{6mygray1}}l}{BET-HiLL}  &\multicolumn{1}{>{\columncolor{6mygray1}}l}{${46.31}$}	&\multicolumn{1}{>{\columncolor{6mygray1}}l}{${40.65}$}	&\multicolumn{1}{>{\columncolor{6mygray1}}l}{${34.99}$}	&\multicolumn{1}{>{\columncolor{6mygray1}}l}{${28.85}$}	&\multicolumn{1}{>{\columncolor{6mygray3}}l}{$22.78$} \\
                            &\multicolumn{1}{>{\columncolor{6mygray2}}l}{DCDT-HiLL}     &\multicolumn{1}{>{\columncolor{6mygray2}}l}{$46.22$}	&\multicolumn{1}{>{\columncolor{6mygray2}}l}{$40.60$}	&\multicolumn{1}{>{\columncolor{6mygray2}}l}{$34.95$}	&\multicolumn{1}{>{\columncolor{6mygray2}}l}{$28.53$}	&\multicolumn{1}{>{\columncolor{6mygray1}}l}{${23.98}$} \\
\toprule[1pt]
\end{tabular}
\end{table}

When it comes to the steganalyzer CC-JRM, both BET-HiLL and the proposed DCDT-HiLL are inferior to J-UNIWARD and GUED. Likewise, there may be two reasons that contributed to the degradation of performance. One is the suboptimal parameter setting for CC-JRM. We simulate DCDT-HiLL with its optimal parameter setting for CC-JRM under Q75, i.e., $p$=0.7 (\emph{DCDT-HiLL-pro}), and then its security performance is indeed improved as shown in Table \ref{tab:Security performance_ori_Q75}. The other is that the DCDT-HiLL and BET-HiLL schemes modify too much mid-to-high frequency coefficients than J-UNIWARD and GUED to resist the detection of JPEG phase-aware feature-based steganalyzers, e.g., GFR and SCA-GFR, which would make their embedding traces easier exposed to steganalyzer CC-JRM. For validation, we remove the integral components of CC-JRM, which are sensitive to the changes of the statistics of DCT modes, especially the mid-to-high frequency modes, and the resulting feature is denoted as crop-CC-JRM-17,270D. Subsequently, applying the crop-CC-JRM to detect the tested schemes at 0.4 bpnzAC under Q75 and Q95, and the comparison results are shown in Table \ref{tab:crop-CC-JRM}. It is observed that the security performance improvements of DCDT-HiLL can reach 2.42\% and 2.85\% at Q75 and Q95, respectively. And so is the BET-HiLL. While for J-UNIWARD and GUED, the improvements are relatively much less. Therefore, the newly emerged JPEG phase-aware feature-based steganalyzers, e.g., GFR and SCA-GFR, are not compatible with the conventional JPEG steganalyzer CC-JRM. Considering that both GFR and its selection-channel aware variant SCA-GFR are currently the most powerful hand-craft JPEG steganalyzers, and the proposed DCDT-Hill is tailored for them by inevitably sacrificing the performance against CC-JRM to some extents.

\begin{table}[h]
\renewcommand\arraystretch{1}
\centering
\caption{Classification error probability ${P_E}$ (in \%) of J-UNIWARD, GUED, BET-HiLL and DCDT-HiLL for CC-JRM and crop-CC-JRM at 0.4 bpnzAC under Q75 and Q95. ($\Delta {P_E}$ is the difference of ${P_E}$ between crop-CC-JRM and CC-JRM.)}\label{tab:crop-CC-JRM}
\begin{tabular}{ccccccc}
\toprule[1pt]
 \multirow{2}{*}{Scheme} & \multicolumn{2}{c}{CC-JRM} & \multicolumn{2}{c}{crop-CC-JRM} & \multicolumn{2}{c}{$\Delta {P_E}$}\\
\cmidrule(lr){2-7}
&Q75 &Q95 &Q75 &Q95 &Q75 &Q95\\
\midrule[0.8pt]
J-UNIWARD 	&$26.83$   &$40.76$   &$26.98$	&$41.36$  &$+0.15$	&$+0.60$\\
\midrule[0.8pt]
GUED 	&$27.18$   &$42.15$   &$28.47$	&$43.29$  &$+1.29$	&$+1.14$\\
\midrule[0.8pt]
BET-HiLL    &$24.74$   &$40.56$   &$27.37$	&$43.32$  &$\bf{+2.63}$	&$\bf{+2.76}$\\
\midrule[0.8pt]
DCDT-HiLL    &$23.39$   &$40.12$   &$25.81$	&$42.97$  &$\bf{+2.42}$	&$\bf{+2.85}$\\
\bottomrule[1pt]
\end{tabular}
\end{table}

\subsection{Practical evaluation of computational complexity}
In this subsection, we further evaluate the computational complexity of our proposed DCDT-HiLL compared to UERD, J-UNIWARD, GUED, and BET-HiLL in terms of computation time (CmpTime). Considering that all the involved JPEG steganographic schemes are implemented under the same framework of STC-based minimal distortion embedding, i.e., the computation of embedding cost for each quantized DCT coefficient + STC encoding, therefore the major difference among them lies in the adopted distortion cost function. And it is quite reasonable to evaluate the computational complexity of the tested schemes by comparing the practical computation times in the calculation of their distortion costs. In our experiment, we calculate the average CmpTimes of the distortion costs for UERD, J-UNIWARD, GUED, BET-HiLL, and DCDT-HiLL, over 2,000 JPEG images randomly selected from BOSSBaseJ75 and BOSSBaseJ95, respectively, using MATLAB 8.2 on a 3.0 GHz Intel Core i5-7400 CPU with 8GB memory. The results are summarized in Table \ref{tab:time_consuming}. It is observed that: 1) the proposed DCDT-HiLL is extremely time-efficient, its CmpTime is one, two, and three orders of magnitude lower than BET-HiLL, GUED, and J-UNIWARD, respectively; 2) DCDT-HiLL could be implemented in a quite affordable time cost as UERD for practical applications.

\begin{table}[h]
\renewcommand\arraystretch{1}
\centering
\caption{Average CmpTimes on a 3.0 GHz Intel Core i5-7400 CPU with 8GB memory over 2,000 JPEG images of $512 \times 512 \times 8$  bits under Q75 and Q95 in calculation of distortion costs for UERD, J-UNIWARD, GUED, BET-HiLL (0.4bpnzAC) and DCDT-HiLL. The unit of time is second (s).}\label{tab:time_consuming}
\begin{tabular}{cccccc}
  \toprule[1pt]
    \centering
    \multirow{2}{*}{QF} & \multicolumn{5}{c}{Average computation times (s)}\\
     \cmidrule(lr){2-6}
     \centering
     &UERD &J-UNIWARD &GUED &BET-HiLL &DCDT-HiLL\\
     \midrule[0.8pt]
      \centering
       75 &$\bm{0.046}$ &$12.12$ &$1.28$ &$0.789$ & $\bm{0.054}$\\
      \midrule[0.8pt]
      \centering
       95 & $\bm{0.051}$ & $12.04$ & $1.29$ & $0.906$& $\bm{0.053}$ \\
  \bottomrule[1pt]
\end{tabular}
\end{table}

\subsection{Further study on the applicability of our proposed scheme}
Recalling the optimal exponent parameter $p$ in the proposed distortion function is obtained experimentally from the specific image database BOSSBase ver1.01 \cite{bas2011break} at Q75 and Q95, therefore the applicability of our proposed scheme for other image database and QFs remains to be further investigated.

\begin{itemize}
\item{Performance of the proposed DCDT-HiLL on other image database}

\ \quad We use image database BOWS2 \cite{BOWS2} to evaluate the applicability of our proposed scheme with the exponent parameter $p$ trained on BOSSBase. For brevity, we only compare the empirical security performance of the proposed DCDT-HiLL with J-UNIWARD, which is one of the most popular JPEG steganographic schemes, using the most effective steganalyzer SCA-GFR on BOWS2J75 and BOWS2J95, which are shown in Table \ref{tab:GUED_BOWS2}. Likewise, the proposed DCDT-HiLL shows an overall superior performance than J-UNIWARD as done in BOSSBase, indicating the effectiveness of our proposed DCDT-HiLL on various databases.

\begin{table*}[h]
\renewcommand\arraystretch{1}
\centering
\caption{Classification error probability ${{P}_E}$ (in \%) of J-UNIWARD and the proposed DCDT-HiLL against steganalyzer SCA-GFR on BOWS2J75 and BOWS2J95.}\label{tab:GUED_BOWS2}
\begin{tabular}{ccccccc}
\toprule[1pt]
 \multirow{2}{*}{QF} & \multirow{2}{*}{Scheme} & \multicolumn{5}{c}{Relative payload $\alpha$ (bpnzAC)} \\
  \cmidrule(lr){3-7}
  \centering
  & &0.1 &0.2 &0.3 &0.4 &0.5\\
\midrule[0.8pt]
\centering
\multirow{2}{*}{75} &J-UNIWARD      &${37.94}$	&$25.06$	&$15.49$	&$8.77$   &$4.59$  \\
                    &{DCDT-HiLL}  &$\bm{38.60}$ 	&$\bm{26.37}$	&$\bm{16.45}$	&$\bm{9.47}$	&$\bm{5.20}$ \\
\midrule[0.8pt]
\centering
\multirow{2}{*}{95} &J-UNIWARD      &$47.04$	&$42.14$	&$35.30$	&$28.05$    &$21.40$ \\
                    &{DCDT-HiLL}  &$\bm{47.06}$ 	&$\bm{42.18}$	&$\bm{35.95}$	&$\bm{29.27}$	&$\bm{22.08}$\\
\bottomrule[1pt]
\end{tabular}
\end{table*}

\item{Performance of the proposed DCDT-HiLL on other QFs}

\ \quad In section 3.2, only the optimal exponent parameters $p$ in the proposed DCDT-HiLL for Q75 and Q95 are investigated, while for other QFs, the empirical rule to determine the corresponding $p$ should be developed, because it is impractical to search for the optimal $p$ for each QF. Note that SCA-GFR is the most effective JPEG steganalyzer and the performance of the proposed DCDT-HiLL with the same parameter setting as SCA-GFR's for other steganalyzers won't change much, then referring to the procedure of determination on the optimal $p$ in section 3.2, we can easily obtain the optimal parameters $p$ for DCDT-HiLL at Q80, Q85, and Q90 in resisting the detection of SCA-GFR as shown in Table \ref{tab:optimal_p_3}.
\begin{table}[h]
\renewcommand\arraystretch{1}
\centering
\caption{The optimal parameter $p$ in DCDT-HiLL for the most effective steganalyzer SCA-GFR at Q75, Q80, Q85, Q90 and Q95.}\label{tab:optimal_p_3}
\begin{tabular}{cccccc}
  \toprule[1pt]
    \multirow{2}{*}{Steganalyzer} & \multicolumn{5}{c}{QF}\\
     \cmidrule(lr){2-6}
     \centering
     &75 &80 &85 &90 &95\\
    \midrule[0.8pt]
        SCA-GFR &$0.5$	&$0.6$ &$0.6$ &$0.8$ &$0.9$\\
    \bottomrule[1pt]
\end{tabular}
\end{table}

\ \quad Then, we can build an empirical rule for parameter $p$ by using an linear regression model w.r.t. $p$ and QF according to the results in Table \ref{tab:optimal_p_3}, i.e.,
\begin{equation}\label{eq:linear_regression_p}
p = 0.02 \times (\rm{QF} - 75) + 0.48.
\end{equation}
The QF in Eq. \eqref{eq:linear_regression_p} is kept in the interval [75, 95]{\footnote{The reason for the selection of interval [75, 95] is that the QFs in this interval are most commonly used in our lives.}}, and as for the one outside this interval, we can follow this procedure and rebuild a new regression model as well. Subsequently, we further compare the empirical security performance of our proposed DCDT-HiLL with J-UNIWARD for steganalyzer SCA-GFR on BOSSBaseJ80, BOSSBaseJ85, and BOSSBaseJ90 using this empirical rule. Referring to the results in Table \ref{tab:GUED_QF}, it is observed that on various QFs, our proposed DCDT-HiLL exhibits better performance than J-UNIWARD as well.

\end{itemize}

\begin{table*}[h]
\renewcommand\arraystretch{1}
\centering
\caption{Classification error probability ${{P}_E}$ (in \%) of J-UNIWARD and the proposed DCDT-HiLL against steganalyzer SCA-GFR on BOSSBaseJ80, BOSSBaseJ85 and BOSSBaseJ90.}\label{tab:GUED_QF}
\begin{tabular}{ccccccc}
\toprule[1pt]
 \multirow{2}{*}{QF} & \multirow{2}{*}{Scheme} & \multicolumn{5}{c}{Relative payload $\alpha$ (bpnzAC)} \\
  \cmidrule(lr){3-7}
  \centering
  & &0.1 &0.2 &0.3 &0.4 &0.5\\
\midrule[0.8pt]
\centering
\multirow{2}{*}{80} &J-UNIWARD      &$38.06$	&$25.88$	&$16.71$	&$10.14$   &$5.92$  \\
                    &{DCDT-HiLL}  &$\bm{38.25}$ 	&$\bm{26.57}$	&$\bm{17.57}$	&$\bm{11.14}$	&$\bm{6.99}$ \\
\midrule[0.8pt]
\centering
\multirow{2}{*}{85} &J-UNIWARD      &$39.41$	&$28.16$	&$19.10$	&$12.15$   &$7.51$  \\
                    &{DCDT-HiLL}  &$\bm{39.57}$ 	&$\bm{28.89}$	&$\bm{20.01}$	&$\bm{13.58}$	&$\bm{8.84}$ \\
\midrule[0.8pt]
\centering
\multirow{2}{*}{90} &J-UNIWARD      &$\bm{42.72}$	&$33.31$	&$24.44$	&$17.33$   &$11.85$  \\
                    &{DCDT-HiLL}  &${42.26}$ 	&$\bm{33.38}$	&$\bm{25.48}$	&$\bm{18.69}$	&$\bm{13.09}$ \\
\bottomrule[1pt]
\end{tabular}
\end{table*}

\subsection{Evaluation on the mutually dependent embedding extension of our proposed scheme}
To verify the claim in section 3.4 that the mutually dependent embedding of our proposed DCDT-HiLL helps to improve the performance, espacilly at Q75, we then compare the performance of the mutually dependent version DCDT-HiLL\_ud with the original DCDT-HiLL on BOSSBaseJ75 at 0.2 bpnzAC against the detection of CC-JRM, GFR and SCA-GFR with their corresponding optimal parameter $p$ setting. Since the computational complexity of DCDT-HiLL\_ud is exponentially increased with $n_k$, we make a constraint that if $n_k$ is large than a threshold $T$, then this block will be skipped for distortion cost updating. In this paper, we set the threshold $T$ and penalty factor $v$ in Eq. \eqref{eq:updating1} as 10, and the results are summarized in Table \ref{tab:non_additive}. It is observed that the performance of the proposed DCDT-HiLL is indeed improved by incorporating the mutually dependent embedding strategy.

\begin{table}[h]
\renewcommand\arraystretch{1}
\centering
\caption{Classification error probability ${{P}_E}$ (in \%) of DCDT-HiLL and DCDT-HiLL\_ud for CC-JRM, GFR and SCA-GFR on BOSSBaseJ75 at 0.2 bpnzAC.}\label{tab:non_additive}
\begin{tabular}{cccc}
  \toprule[1pt]
  \centering
  \multirow{2}{*}{Scheme} & \multicolumn{3}{c}{Steganalyzer}\\
  \cmidrule(lr){2-4}
    \centering
    &CC-JRM &GFR &SCA-GFR\\
  \midrule[0.8pt]
  \centering
   DCDT-HiLL  &$40.10$  &$29.95$ &$24.51$\\
  \midrule[0.8pt]
  \centering
   DCDT-HiLL\_ud &$\bm{40.46}$  &$\bm{30.69}$  &$\bm{25.30}$\\
  \bottomrule[1pt]
\end{tabular}
\end{table}

\section{Conclusion}
In this paper, a novel Distortion Cost Domain Transformation (DCDT) based JPEG steganographic scheme is proposed, which formulates the JPEG steganography as the optimization problem of minimizing the overall distortion cost in its decompressed spatial domain, aiming to maintain the statistical undetectability in both spatial and DCT domains. The proposed DCDT scheme transforms the decompressed $8 \times 8$ spatial pixel block distortion costs into DCT domain by incorporating a generalized domain distortion cost transformation function in terms of the embedding changes in decompressed $8\times8$ pixel block and the adopted distortion cost function in spatial domain. The domain distortion cost transformation function is developed with an exponential model to further maintain the statistical undetectability in both spatial and JPEG domains. Extensive experiments have been carried out, which demonstrates that the proposed DCDT-Hill outperforms other existing SOTA JPEG stgeanographic schemes, including UERD, J-UNIWARD, and GUED, in resisting the detection of newly emerged phase-aware JPEG steganalyzers, e.g., GFR and SCA-GFR. In addition, the proposed DCDT-HiLL can rival the SOTA BET-HiLL with one order of magnitude lower computational complexity as well. The experimental results also show that our proposed DCDT-HiLL has strong applicability, and its security performance can be further improved by incorporating the mutually dependent embedding strategy. Overall, the proposed DCDT-HiLL can not only improve the performance against JPEG phase-aware feature-based steganalyzers but also broaden the applications of existing image steganographic schemes in spatial domain.

\section*{Acknowledgement}
This work was supported in part by the National Natural Science Foundation of China under Grant U1736215, Grant U1936212, and Grant 61772573.

\section*{References}
\bibliography{mybibfile}

\end{document}